\documentclass{aa}  

\usepackage{graphicx}
\usepackage{xcolor}
\usepackage{subfigure}          % allows pictures to be placed side by side
\usepackage{placeins}
\usepackage{txfonts}
\usepackage{hyperref}           % to add links in your PDF file, use the
\usepackage{acronym}            % manages acronyms

\acrodef{JWST}[JWST]{James Webb Space Telescope}
\acrodef{MHD}[MHD]{magnetohydrodynamic}
\acrodef{MIRI}[MIRI]{Mid-Infrared Instrument}
\acrodef{MINDS}[MINDS]{MIRI Mid-Infrared Disk Survey}

\begin{document}

\title{Changing disc compositions via internal photoevaporation I: Solar-mass stars}

\author{J. L. Lienert\inst{1}\fnmsep\thanks{Corresponding author; \email{lienert@mpia.de}}, B. Bitsch\inst{1,2} \and Th. Henning\inst{1}}

\institute{Max-Planck-Institut für Astronomie, Königstuhl 17, D-69117 Heidelberg, Germany \and University College Cork, College Rd, Cork T12 K8AF, Ireland}

\date{Received November 30, 2023; accepted February 13, 2024}

\abstract{The chemical evolution of protoplanetary discs is a complex process that is not fully understood. Several factors influence the final spatial distribution of atoms and molecules in the disc. One such factor is the inward drift and evaporation of volatile-rich pebbles that can enrich the inner disc with vapour. In particular, the inner disc is first enriched with evaporating water-ice, resulting in a low C/O ratio, before carbon-rich gas from the outer disc ---originating from the evaporation of CO, CO$_2$, and CH$_4$ ice--- is transported viscously inwards, elevating the C/O ratio again. However, it is unclear how internal photoevaporation ---which carries away gas and opens gaps in the disc that can block inward drifting pebbles--- affects the chemical composition of the disc. Our goal is to study how and to what extent internal photoevaporation and the subsequent opening of gaps influence the chemical evolution of protoplanetary discs around solar-like stars ($M_{\star} = 1 \, M_{\odot}$), where we specifically focus on the C/O ratio and the water content. To carry out our simulations, we use a semi-analytical 1D disc model. The code \texttt{chemcomp} includes viscous evolution and heating, pebble growth and drift, pebble evaporation and condensation, as well as a simple chemical partitioning model for the disc. We show that internal photoevaporation plays a major role in the evolution of protoplanetary discs and their chemical composition: As photoevaporation opens a gap, inward drifting pebbles are stopped and can no longer contribute to the volatile content in the gas. In addition, volatile-rich gas from the outer disc, originating from evaporated CO, CO$_2$, or CH$_4$ ice, is carried away by the photoevaporative winds. Consequently, the C/O ratio in the inner disc remains low. In contrast, gaps opened by giant planets still allow the gas to pass, resulting in an elevated C/O ratio in the inner disc, similar to the evolution of viscous discs without internal photoevaporation. This opens the possibility to distinguish observationally between these two scenarios when measuring the C/O ratio, implying that we can infer the root cause of deep gap structures when observing protoplanetary discs. In the case of a clear separation of the disc by photoevaporation, we additionally find an elevated water content in the inner disc, because the water vapour and ice undergo a cycle of evaporation and recondensation, preventing the inward accretion of water onto the star, in contrast to the situation for hydrogen and helium. We conclude that it is very difficult to achieve supersolar C/O ratios in the inner parts of protoplanetary discs when taking internal photoevaporation into account. This indicates the potential importance of photoevaporation for understanding the chemical evolution of these discs and the planets forming in them.}


\maketitle

%-------------------------------------------------------------------%

\section{Introduction} \label{sec:Introduction}
A large number of exoplanets have now been characterised, showing very diverse planetary system architectures and exoplanet properties. However, their exact formation pathway and composition are not fully understood \citep[for a review, see][]{drazkowskaPlanetFormationTheory2022}. It is therefore important to study their formation environment via simulations in order to gain greater insight into the structure and composition of protoplanetary discs and their time evolution. This in turn helps us to constrain the properties and formation scenarios of planets. \\
In the core accretion scenario, the planetary core forms via solid accretion. This solid accretion can come in the form of planetesimals \citep{idaDeterministicModelPlanetary2004,guileraPlanetesimalFragmentationGiant2014,mordasiniPlanetaryPopulationSynthesis2018,miguelDiverseOutcomesPlanet2019,emsenhuberNewGenerationPlanetary2021} or pebbles \citep{bitschGrowthPlanetsPebble2015,bitschFormationPlanetarySystems2019,nduguPlanetPopulationSynthesis2018,liuSuperEarthMassesSculpted2019,izidoroFormationPlanetarySystems2021}, where pebble accretion is especially more efficient than planetesimal accretion in the outer disc regions \citep[e.g.][]{johansenExploringConditionsForming2019}. Planetesimal accretion models have traditionally been favoured, but models via pebble accretion have recently become popular. Most of these population synthesis studies tried to reproduce only observed planetary masses and orbital distances. However, it is now possible to constrain planet formation models even further with newly measured atmospheric compositions of observed exoplanets (e.g. \citeauthor{molliereRetrievingAtmosphericProperties2020} et al. \citeyear{molliereRetrievingAtmosphericProperties2020}; \citeauthor{lineSolarSubsolarMetallicity2021} \citeyear{lineSolarSubsolarMetallicity2021}; \citeauthor{pelletierWhereWaterJupiterlike2021} \citeyear{pelletierWhereWaterJupiterlike2021}; \citeauthor{molliereInterpretingAtmosphericComposition2022} \citeyear{molliereInterpretingAtmosphericComposition2022}; \citeauthor{augustConfirmationSubsolarMetallicity2023} \citeyear{augustConfirmationSubsolarMetallicity2023}). Recent theoretical studies from \cite{mordasiniIMPRINTEXOPLANETFORMATION2016}, \cite{schneiderHowDriftingEvaporating2021,schneiderHowDriftingEvaporating2021a} or \cite{bitschHowDriftingEvaporating2022} tried to additionally incorporate these atmospheric abundances. \\
Planetary compositions originate from the composition of the disc they form in. Therefore, it is very important to simulate the full evolution of the structure and chemical composition of protoplanetary discs. The latter is determined by the disc temperature and density, radiation fields, and the position of evaporation lines of different species \citep[see e.g.][]{obergEFFECTSSNOWLINESPLANETARY2011,henningChemistryProtoplanetaryDisks2013,schneiderHowDriftingEvaporating2021,eistrupChemicalEvolutionIces2022,molliereInterpretingAtmosphericComposition2022}. \\
Even if the temperature of the disc were to remain constant over time, the chemical composition of the disc would still evolve. Pebbles drifting through the disc evaporate their volatile content at the evaporation lines, therefore altering its composition \citep[e.g.][]{pisoSNOWLINELOCATIONSPROTOPLANETARY2015,boothChemicalEnrichmentGiant2017,schneiderHowDriftingEvaporating2021,kalyaanEffectDustEvolution2023}. On the other hand, chemical reactions play only a minor role in the chemical evolution of pebbles in discs because their drift timescales are much shorter than those of the chemical reactions \citep{boothPlanetformingMaterialProtoplanetary2019,eistrupChemicalEvolutionIces2022}. \\
The composition and evolution of the disc are influenced by pressure bumps that can block inward drifting pebbles \citep[e.g.][]{pinillaTrappingDustParticles2012}. Such traps can be caused by planets reaching their pebble isolation mass and creating gaps in the discs \citep[e.g.][]{paardekooperDustFlowGas2006,lambrechtsSeparatingGasgiantIcegiant2014,ataieeHowMuchDoes2018,bitschPebbleisolationMassScaling2018}. These gaps influence the chemical composition of inner discs \citep[see e.g.][]{bitschDryWaterWorld2021,schneiderHowDriftingEvaporating2021,schneiderHowDriftingEvaporating2021a,kalyaanEffectDustEvolution2023}. \cite{banzattiJWSTRevealsExcess2023}, \cite{grantMINDSDetection132023}, \cite{perottiWaterTerrestrialPlanetforming2023}, and \cite{taboneRichHydrocarbonChemistry2023} studied the chemical composition of discs via \ac{JWST} observations, showing that pressure bumps can indeed influence the chemical composition of inner discs. \\
However, disc evolution is not only determined by viscous effects. Evaporation via \ac{MHD} winds \citep[for a review, see][]{lesurHydroMagnetohydroDustGas2023} or photoevaporation \citep[for a review, see][]{pascucciRoleDiskWinds2022} plays a significant role as well because these processes carry away disc material, which, in the case of photoevaporation, carves deep gaps into the disc structure. Thus, these effects not only influence disc evolution and their late stages of dispersal but also change the disc's chemical composition. Photoevaporation becomes especially important as it determines the final stages of disc evolution \citep[e.g.][]{ercolanoXRAYIRRADIATEDPROTOPLANETARY2009,pascucciEVIDENCEDISKPHOTOEVAPORATION2009,ercolanoMetallicityPlanetFormation2010,owenTheoryDiscPhotoevaporation2012a,owenCharacterizingThermalSweeping2013}. Such a disc dissolving process is necessary since it is known from observations that discs only live for a few million years \citep{mamajekInitialConditionsPlanet2009,fedeleTimescaleMassAccretion2010}, even though the disc lifetime can vary significantly depending on the stellar type of the system \citep[e.g.][]{michelBridgingGapProtoplanetary2021,pfalznerMostPlanetsMight2022}. \\
Photoevaporation is a process in which high-energy radiation transfers so much of its energy to gas particles in the disc that their velocities become high enough to escape. Energetic radiation can come from nearby stars outside the system, in which case the evaporation process is referred to as external photoevaporation \citep[for a review, see][]{winterExternalPhotoevaporationPlanetforming2022}. Internal photoevaporation on the other hand is where high-energy radiation from the host star is used to energise particles in the disc \citep[for a review, see][]{pascucciRoleDiskWinds2022}. Gaps created by photoevaporation block inward-moving pebbles. At the same time, gas is carried away by photoevaporative winds, thus cutting off the inner regions of the disc from their supply from outer areas. \\
In this work, we combine a viscous disc evolution model ---that includes pebble drift and evaporation--- with a model of internal photoevaporation. We study protoplanetary discs around solar-like stars ($M_{\star} = 1 \, M_{\odot}$) and compare our results first to a purely viscous disc and second to a disc with a forming giant planet to distinguish between the two kinds of gaps that form for each case. \\
This paper is structured as follows: Section \ref{sec:Methods} includes an outline of our model and an explanation of the numerical methods used in \texttt{chemcomp}. The obtained results are shown in section \ref{sec:Results}, and their implications are discussed in section \ref{sec:Discussion} before we conclude with section \ref{sec:Summary_and_conclusions}.

\section{Methods} \label{sec:Methods}
\begin{table*}
    \centering
        \begin{tabular}{ccc}
            \hline
            Species (Y)         & $T_{\text{cond}}$ [K] & $v_{\text{Y}}$                    \\ \hline \hline
            CO                  & $20$                  & $0.45 \: \times \: \text{C/H}$    \\
            N$_2$               & $20$                  & $0.45 \: \times \: \text{N/H}$    \\
            CH$_4$              & $30$                  & $0.45 \: \times \: \text{C/H}$    \\
            CO$_2$              & $70$                  & $0.1 \: \times \: \text{C/H}$     \\
            NH$_3$              & $90$                  & $0.1 \: \times \: \text{N/H}$     \\
            H$_2$S              & $150$                 & $0.1 \: \times \: \text{S/H}$     \\
            H$_2$O              & $150$                 & $\text{O/H} - \left( 3 \times \text{MgSiO}_3/\text{H} + 4 \times \text{Mg}_2\text{SiO}_4/\text{H} + \text{CO/H} + 2 \times \text{CO}_2/\text{H} + 3 \times \text{Fe}_2\text{O}_3/\text{H} \right.$    \\
                                &                       & $ \left. + \: 4 \times \text{Fe}_3\text{O}_4/\text{H} + \text{VO/H} + \text{TiO/H} + 3 \times \text{Al}_2\text{O}_3 + 8 \times \text{NaAlSi}_3\text{O}_8 + 8 \times \text{KAlSi}_3\text{O}_8 \right)$ \\
            Fe$_3$O$_4$         & $371$                 & $(1/6) \times (\text{Fe/H} - 0.9 \times \text{S/H})$                          \\
            FeS                 & $704$                 & $0.9 \: \times \: \text{S/H}$     \\
            NaAlSi$_3$O$_8$     & $958$                 & Na/H                              \\
            KAlSi$_3$O$_8$      & $1006$                & K/H                               \\
            Mg$_2$SiO$_4$       & $1354$                & $\text{Mg/H} - (\text{Si/H} - 3 \times \text{K/H} - 3 \times \text{Na/H})$    \\
            Fe$_2$O$_3$         & $1357$                & $0.25 \times (\text{Fe/H} - 0.9 \times \text{S/H})$                           \\
            VO                  & $1423$                & V/H                               \\
            MgSiO$_3$           & $1500$                & $\text{Mg/H} - 2 \times (\text{Mg/H} - (\text{Si/H} - 3 \times \text{K/H} - 3 \times \text{Na/H}))$   \\
            Al$_2$O$_3$         & $1653$                & $0.5 \times (\text{Al/H} - (\text{K/H} + \text{Na/H}))$                       \\
            TiO                 & $2000$                & Ti/H                              \\ \hline
        \end{tabular}
    \caption{Condensation temperatures and volume mixing ratios of the chemical species in our model. The condensation temperatures are taken from \cite{loddersSolarSystemAbundances2003}. We assume constant $T_{\text{cond}}$ as its dependence on pressure is marginal \citep{mousisDETERMINATIONMINIMUMMASSES2009}. For simplification, we treat condensation and sublimation temperature as the same. Note that for Fe$_2$O$_3$, the condensation temperature for pure iron is adopted \citep{loddersSolarSystemAbundances2003}. The $v_{\text{Y}}$'s (i.e. by number) are adopted for the species as a function of disc elemental abundances \citep[e.g.][]{madhusudhanExoplanetaryAtmospheres2014}; they are expanded with the different mixing ratios from \cite{bitschInfluenceSubSupersolar2020}.}
    \label{table:molecular_species}
\end{table*}

\begin{table*}
    \centering
        \begin{tabular}{ccccccccc}
            \hline
            $M_{\star} [M_{\odot}]$ & $a$       & $b$       & $c$           & $d$       & $e$           & $f$       & $g$           & $\dot{M}_{\text{w}} [10^{-8} \, M_{\odot} \text{yr}^{-1}]$    \\ \hline \hline
            $1.0$                   & $-0.6344$ & $6.3587$  & $-26.1445$    & $56.4477$ & $-67.7403$    & $43.9212$ & $-13.2316$    & $3.86446$                                                     \\ \hline
        \end{tabular}
    \caption{Fit parameters for the surface density mass loss profile taken from \cite{picognaDispersalProtoplanetaryDiscs2021a}.}
    \label{table:fit_parameters}
\end{table*}

\begin{figure}
    \centering
    \begin{minipage}{0.5\textwidth}
        \includegraphics[width=\textwidth]{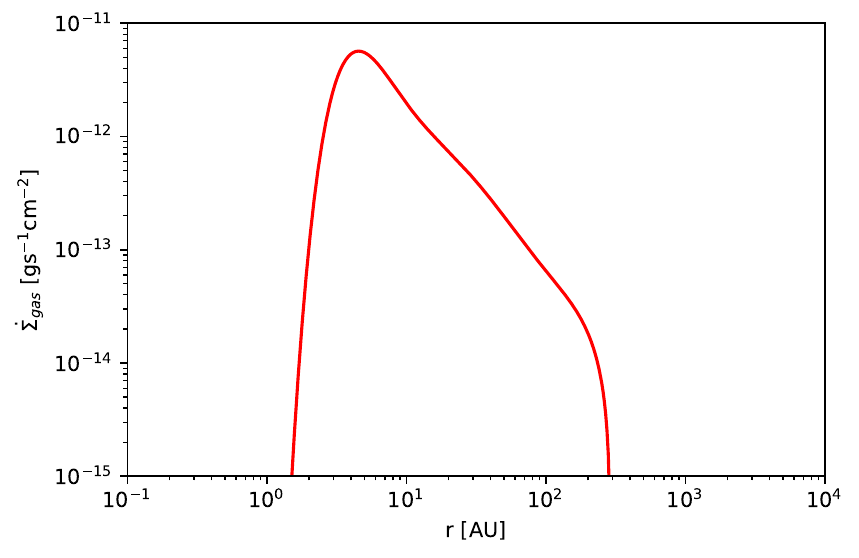}
    \end{minipage}
    \caption{Gas surface density loss rate as a function of disc radius, as given in equation \ref{eq:sig_dot_ph}, taken from \cite{picognaDispersalProtoplanetaryDiscs2021a}.}
    \label{fig:sig_dot}
\end{figure}

\begin{table}
    \centering
        \begin{tabular}{cc}
            \hline
            Species (X) & Abundance             \\ \hline \hline
            He/H        & $0.085$               \\
            O/H         & $4.90 \cdot 10^{-4}$  \\
            C/H         & $2.69 \cdot 10^{-4}$  \\
            N/H         & $6.76 \cdot 10^{-5}$  \\
            Mg/H        & $3.98 \cdot 10^{-5}$  \\
            Si/H        & $3.24 \cdot 10^{-5}$  \\
            Fe/H        & $3.16 \cdot 10^{-5}$  \\
            S/H         & $1.32 \cdot 10^{-5}$  \\
            Al/H        & $2.82 \cdot 10^{-6}$  \\
            Na/H        & $1.74 \cdot 10^{-6}$  \\
            K/H         & $1.07 \cdot 10^{-7}$  \\
            Ti/H        & $8.91 \cdot 10^{-8}$  \\
            V/H         & $8.59 \cdot 10^{-9}$  \\ \hline
        \end{tabular}
    \caption{Number ratios of the elements used in our model, corresponding to the abundance of element X relative to that of hydrogen in the solar photosphere \citep{asplundChemicalCompositionSun2009}.}
    \label{table:initial_abundances}
\end{table}

\begin{table}
    \centering
        \begin{tabular}{cc}
            \hline
            Simulation parameter                    & Value                 \\ \hline \hline
            Stellar mass $M_{\star}$                & $1 \, M_{\odot}$      \\
            Stellar luminosity $L_{\star}$          & $1 \, L_{\odot}$      \\
            Viscous parameter $\alpha$              & $10^{-4}$             \\
            Initial disc mass $M_{\text{disc}}$     & $0.1 \, M_{\odot}$    \\
            Initial disc radius $R_{\text{disc}}$   & $75 \, \text{AU}$     \\
            Initial dust-to-gas ratio               & $2 \%$                \\ \hline
        \end{tabular}
    \caption{List of parameters used for our standard simulations.}
    \label{table:sim_par}
\end{table}

For our numerical simulations, we use the \texttt{chemcomp} code as described in \cite{schneiderHowDriftingEvaporating2021}. The code is a 1D, semi-analytical model of protoplanetary discs coupled to a planetary growth model, where the latter is only used for one case where we compare the effects of a planet on the disc structure. The details of the planet formation model can be found in \cite{schneiderHowDriftingEvaporating2021}. The disc evolution follows a classic viscous evolution model \citep[see e.g.][]{lynden-bellEvolutionViscousDiscs1974a}, making use of the alpha-viscosity description from \cite{shakuraBlackHolesBinary1973}. \\
In the disc, dust grains can grow to pebbles, with their growth being modelled by the two-population approach developed by \cite{birnstielSimpleModelEvolution2012}, where the grains exist in two representative sizes, the monomer size and the size of a large grain population. Their growth is being limited by drift, fragmentation and drift-induced fragmentation. The disc is enriched by volatiles evaporating from the inward drifting pebbles at the particular ice lines of each species \citep{schneiderHowDriftingEvaporating2021}.

\subsection{Viscous evolution}
The viscosity is given by
\begin{equation}
    \nu = \alpha \frac{c_{\text{s}}^2}{\Omega_{\text{K}}},
\end{equation}
where $\alpha$ is a dimensionless factor describing the strength of the turbulence, $c_{\text{s}}$ is the isothermal sound speed and $\Omega_{\text{K}} = \sqrt{\frac{GM_{\star}}{r^3}}$ is the Keplerian angular frequency. The speed of sound can be linked to the mid-plane temperature of the disc. The temperature is calculated from an equilibrium between viscous and stellar heating with radiative cooling. In our model, it does not evolve in time for simplicity and is thus calculated only at the initialisation of the simulation. \\
The time evolution of the gas surface density in our disc is given by the viscous disc equation, which can be derived from the conservation of mass and angular momentum \citep{pringleAccretionDiscsAstrophysics1981,armitageAstrophysicsPlanetFormation2013},
\begin{equation} \label{eq:viscous_disc_equation}
    \frac{\partial \Sigma_{\text{gas,Y}}}{\partial t} - \frac{3}{r} \frac{\partial}{\partial r} \left[ \sqrt{r} \frac{\partial}{\partial r} \left( \sqrt{r} \nu \Sigma_{\text{gas,Y}} \right) \right] = \dot{\Sigma}_{\text{Y}},
\end{equation}
where $\dot{\Sigma}_{\text{Y}}$ is the source term of the molecular species Y, as given in table \ref{table:molecular_species}. It originates from the evaporation and condensation of pebbles and is given by
\begin{equation}
    \dot{\Sigma}_{\text{Y}} = \left\{
        \begin{matrix}
            \dot{\Sigma}_{\text{Y}}^{\text{evap}} \quad r <    r_{\text{ice,Y}} \\
            \dot{\Sigma}_{\text{Y}}^{\text{cond}} \quad r \geq r_{\text{ice,Y}}.
        \end{matrix}
    \right.
\end{equation}
Here, $\dot{\Sigma}_{\text{Y}}^{\text{evap}}$ and $\dot{\Sigma}_{\text{Y}}^{\text{cond}}$ are the evaporation and condensation source terms of species Y, originating from the evaporation and condensation of volatiles of species Y at the respective ice line $r_{\text{ice,Y}}$.

\subsection{Disc lifetime and internal photoevaporation}
When a viscous disc is subject to internal photoevaporation, an additional term $\dot{\Sigma}_{\text{w}}$, describing the photoevaporative gas surface density loss rate, is subtracted from the right-hand-side of the disc's viscous equation \ref{eq:viscous_disc_equation},
\begin{equation}
    \frac{\partial \Sigma_{\text{gas,Y}}}{\partial t} - \frac{3}{r} \frac{\partial}{\partial r} \left[ \sqrt{r} \frac{\partial}{\partial r} \left( \sqrt{r} \nu \Sigma_{\text{gas,Y}} \right) \right] = \dot{\Sigma}_{\text{Y}} - \dot{\Sigma}_{\text{w}}.
\end{equation}
We use a description for internal photoevaporation due to X-rays following the work of \cite{picognaDispersalProtoplanetaryDiscs2019a,picognaDispersalProtoplanetaryDiscs2021a} and \cite{ercolanoDispersalProtoplanetaryDiscs2021a}. They only use the soft X-ray regime as it has shown to be the most efficient for internal photoevaporation. The surface density loss is given by
\begin{align} \label{eq:sig_dot_ph}
    \dot{\Sigma}_{\text{w}} &= \left( \frac{6 a \ln (R)^5}{R \ln(10)^6} + \frac{5 b \ln (R)^4}{R \ln(10)^5} + \frac{4 c \ln (R)^3}{R \ln(10)^4} + \frac{3 d \ln (R)^2}{R \ln(10)^3} \right. \\
    &\quad \enspace + \left. \frac{2 e \ln (R)}{R \ln(10)^2} + \frac{f}{R \ln(10)} \right) \frac{\ln(10)}{2 \pi R} \dot{M}_{\text{w}} [M_{\odot} \, \text{AU}^{-2} \, \text{yr}^{-1}], \nonumber
\end{align}
with
\begin{align}
    \frac{\dot{M}_{\text{w}} (R)}{\dot{M}_{\text{w}} (L_\text{X})} = \quad &10^{a \log_{10} (R)^6 + b \log_{10} (R)^5 + c \log_{10} (R)^4 + d \log_{10} (R)^3} \\
    \quad \cdot &10^{e \log_{10} (R)^2 + f \log_{10} (R) + g}, \nonumber
\end{align}
where the fit parameters for a solar-mass star are presented in table \ref{table:fit_parameters}. \\
The loss rate of the gas surface density is visualised in figure \ref{fig:sig_dot} as a function of disc radius. It is constant in time but varies in radius, meaning that the same amount of disc material is taken away in each time step of our simulations. The peak of photoevaporative mass loss is around $5 \, \text{AU}$, with the mass generally being lost between $1.7 \, \text{AU}$ and $119 \, \text{AU}$, enabling the creation of a gap in this area. Disc evolution in regions unaffected by photoevaporation is then purely determined by the viscous transport of gas and dust.

\subsection{Pebble drift and evaporation}
In our model, protoplanetary discs consist of gas and dust grains. The latter can grow to millimetre- to centimetre-sized pebbles, with their growth being limited by drift, turbulent fragmentation and drift-induced fragmentation. Gas and pebbles will move through the disc, with the pebbles drifting faster due to gas drag and therefore reaching the inner disc first. On their way through the disc, the pebbles will cross several evaporation fronts of different molecular species and evaporate their respective volatile content. Enrichment of the inner disc is therefore a natural outcome of our model. Gas close to the host star will eventually be accreted onto it. It is also possible for gaseous species to recondense at such evaporation lines when crossing them in the outwards direction, which can lead to large enhancements of these particular solids compared to the solar composition \citep[e.g.][]{aguichineMassRadiusRelationships2021,mousisSituExplorationAtmospheres2022,mahFormingSuperMercuriesRole2023}. \\
Dust grains move through the disc with a radial velocity $u_{\text{Z}}$ \citep{weidenschillingAerodynamicsSolidBodies1977,brauerCoagulationFragmentationRadial2008,birnstielSimpleModelEvolution2012}
\begin{equation} \label{eq:radial_velocity_dust_grains}
    u_{\text{Z}} = \frac{1}{1 + \text{St}^2} u_{\text{gas}} - \frac{2}{\text{St}^{-1} + \text{St}} \Delta v,
\end{equation}
where $\text{St}$ is the Stokes number of a particle, and $u_{\text{gas}}$ and $\Delta v$ are the velocity of the gas and its azimuthal speed, respectively. The latter is defined by
\begin{equation}
    \Delta v = v_{\text{K}} - v_{\varphi} = - \frac{1}{2} \frac{\text{d} \ln P}{\text{d} \ln r} \left( \frac{H_{\text{gas}}}{r} \right)^2 v_{\text{K}},
\end{equation}
where $v_{\text{K}} = \Omega_{\text{K}} r$ is the Keplerian velocity, $P$ is the gas pressure and $H_{\text{gas}} = \frac{c_{\text{s}}}{\Omega_{\text{K}}}$ is the scale height of the disc. \\
The two-population approach by \cite{birnstielSimpleModelEvolution2012} used here divides the grain size distribution into two groups: small and large grains. This model then takes the average of their velocities. The code self-consistently stops the inward drifting pebbles at pressure bumps. \\
Equation \ref{eq:radial_velocity_dust_grains} implies that small dust grains growing larger increase their radial velocity. On the other hand, their growth can be limited by their velocity. When exceeding a certain velocity limit, the dust grains will fragment when colliding \citep{birnstielDustRetentionProtoplanetary2009}. This fragmentation velocity has been measured in laboratory experiments to values of $1 \, \text{m/s}$ to $10 \, \text{m/s}$ \citep{poppeAnalogousExperimentsStickiness2000,gundlachSTICKINESSMICROMETERSIZEDWATERICE2014}. We use here a variable fragmentation velocity that changes its value from $1 \, \text{m/s}$ to $10 \, \text{m/s}$ at the water-ice line in the radial direction. This is motivated by the idea that water-ice grains can reach larger fragmentation velocities before they fragment compared to silicate grains (\citeauthor{gundlachSTICKINESSMICROMETERSIZEDWATERICE2014} \citeyear{gundlachSTICKINESSMICROMETERSIZEDWATERICE2014}, but see also \citeauthor{musiolikContactsWaterIce2019} \citeyear{musiolikContactsWaterIce2019}).

\subsection{Chemistry model and initial conditions} \label{ssec:Initial_conditions}
We assume that the disc's original chemical composition is similar to the one of its host star. Since we investigate solar-like stars, it is reasonable to use solar abundances [X/H] for our initial disc composition, see table \ref{table:initial_abundances} for the exact values which are taken from \cite{asplundChemicalCompositionSun2009}. Since the disc temperature varies according to the orbital distance from the star, the disc composition is likewise dependent on the disc radius. \\
Elements are distributed into molecules Y following a simple chemical partitioning model from \cite{schneiderHowDriftingEvaporating2021}, based on \cite{bitschInfluenceSubSupersolar2020}. Molecules from a certain species Y are then either present in gaseous or solid form, dependent on their condensation temperature and position in the disc. If they are at a location where the disc temperature is higher than their condensation temperature, the species are available as vapour, and if they are at a location where the disc temperature is lower than their condensation temperature, the species are available in condensed form. The point where the mid-plane temperature is equal to the condensation temperature of a certain molecule species is the evaporation line of that species. \\
The number ratio of two elemental species $\text{X}_1$ and $\text{X}_2$ is defined via
\begin{equation}
    \text{X}_1 / \text{X}_2 = \frac{m_{\text{X}_1}}{m_{\text{X}_2}} \frac{\mu_{\text{X}_2}}{\mu_{\text{X}_1}},
\end{equation}
where $m_{\text{X}_1}$ and $m_{\text{X}_2}$ are the mass fractions of the two elements in a medium of mass $m$ (e.g. $m = \Sigma_{\text{gas}}$) and $\mu_{\text{X}_1}$ and $\mu_{\text{X}_2}$ are the atomic masses of the specific elements. In this paper, this definition is used to calculate the C/O ratio. \\
For our standard simulations, we use a star with a mass of $M_{\star} = 1 \, M_{\odot}$ and a luminosity of $L_{\star} = 1 \, L_{\odot}$. Our disc is determined by a viscous $\alpha$ parameter of $\alpha = 10^{-4}$, an initial disc mass of $M_{\text{disc}} = 0.1 \, M_{\odot}$, an initial disc radius of $R_{\text{disc}} = 75 \, \text{AU}$ and an initial dust-to-gas ratio of $2 \%$, as listed in table \ref{table:sim_par}. Our initial gas surface density profile is determined by an exponential decay in the outer disc regions. The initial radius is the point where the exponential cut-off sets in.

\section{Results} \label{sec:Results}
\begin{figure*}
    \centering
    \begin{minipage}{\textwidth}
        \subfigure{\includegraphics[width=0.5\textwidth]{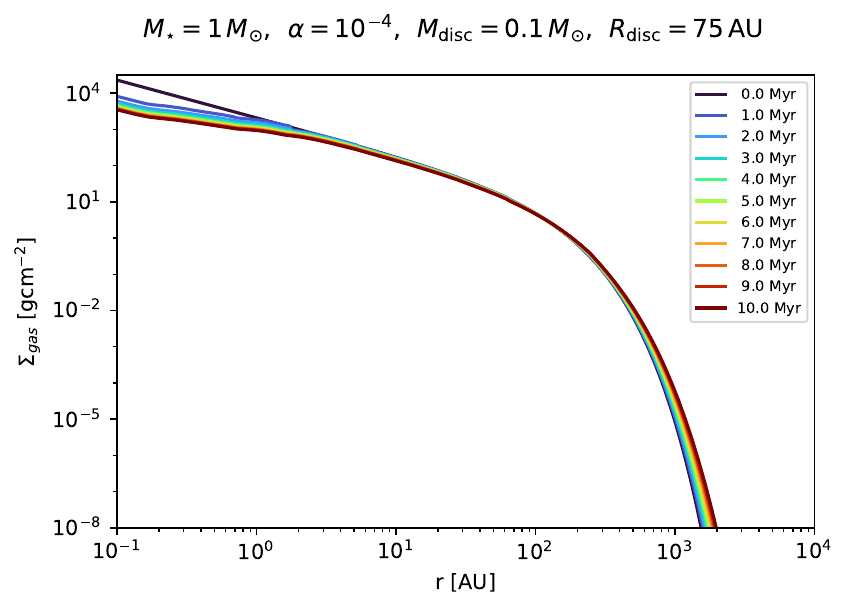}}
        \subfigure{\includegraphics[width=0.5\textwidth]{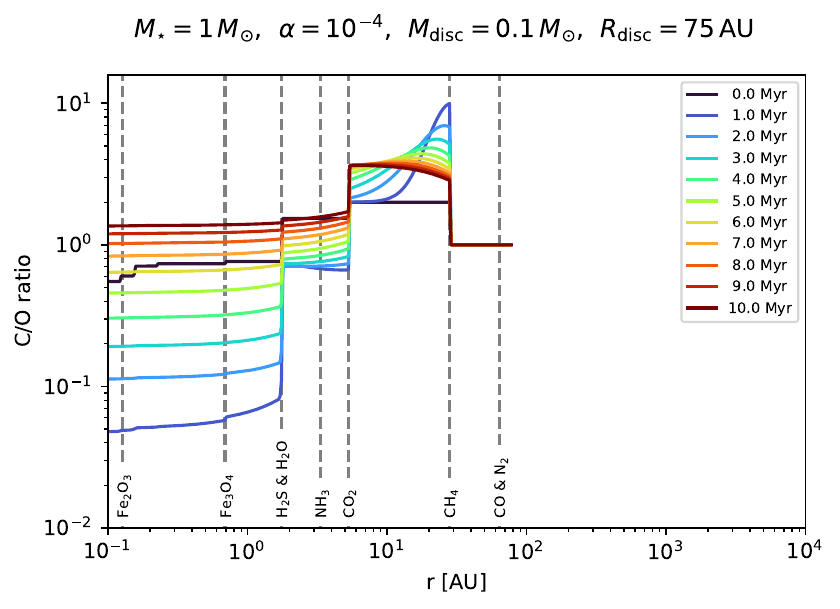}}
        \subfigure{\includegraphics[width=0.5\textwidth]{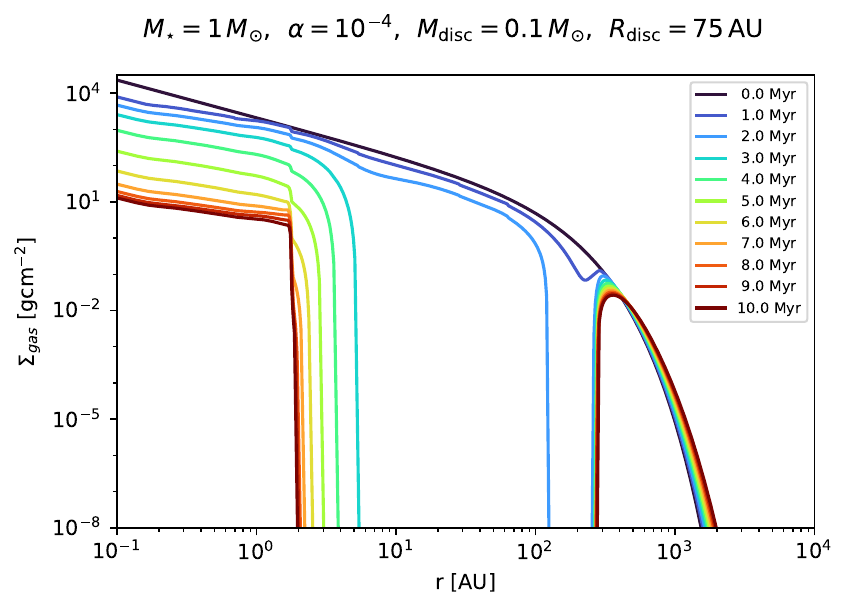}}
        \subfigure{\includegraphics[width=0.5\textwidth]{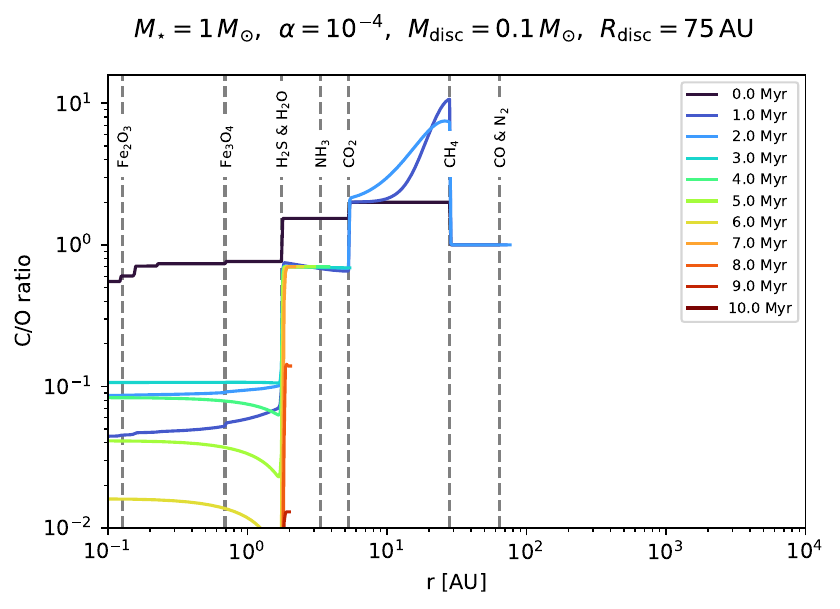}}
        \subfigure{\includegraphics[width=0.5\textwidth]{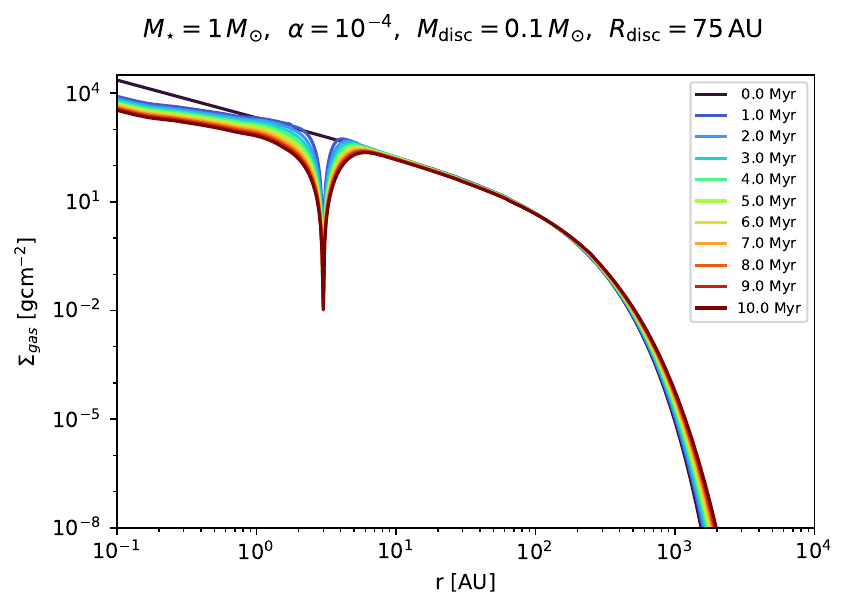}}
        \subfigure{\includegraphics[width=0.5\textwidth]{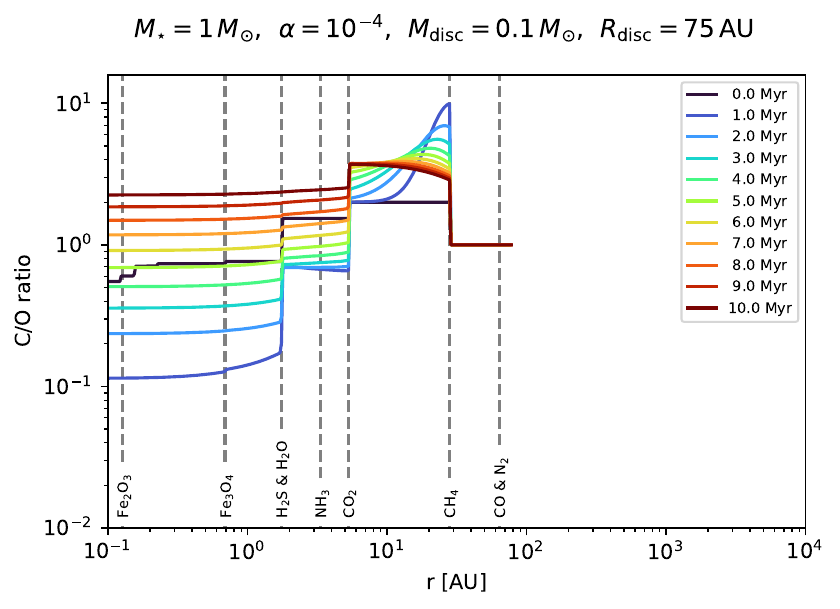}}
    \end{minipage}
    \caption{Disc evolution for different disc configurations. \textbf{Top:} Viscous disc without internal photoevaporation. \textbf{Middle:} Viscous disc with internal photoevaporation due to X-rays. \textbf{Bottom:} Viscous disc without internal photoevaporation where a planet seed is put at $3 \, \text{AU}$ at $0.05 \, \text{Myr}$. The planet has a final mass of about $2860 \, M_{\text{Earth}}$ and reaches pebble isolation mass at $0.1 \, \text{Myr}$. \textbf{Left:} Gas surface density as a function of disc radius, time evolution is shown in colour - from black, which corresponds to $0 \, \text{Myr}$, to dark red, which corresponds to $10 \, \text{Myr}$. \textbf{Right:} Gaseous C/O ratio as a function of disc radius and time (colour-coded). The evaporation lines for different molecules are given as grey dashed lines. Note that the C/O ratio is calculated from number densities scaled to the solar value and that we, by definition, have no specified C/O ratio in the gas phase beyond the CO evaporation front. We use our standard parameters for this simulation, as given in table \ref{table:sim_par}.}
    \label{fig:gas_surface_density_and_C_to_O_no_ph}
\end{figure*}

\begin{figure}
    \centering
    \begin{minipage}{0.5\textwidth}
        \includegraphics[width=\textwidth]{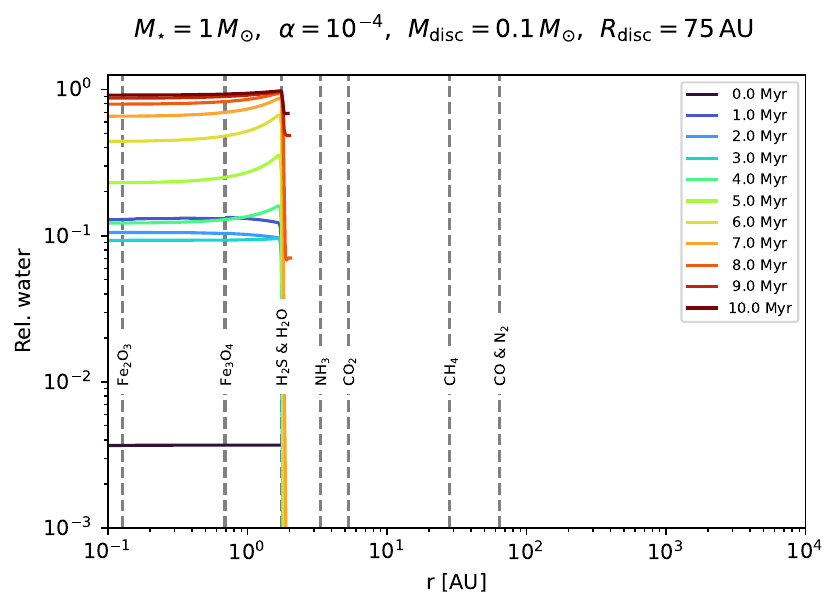}
    \end{minipage}
    \caption{Relative water content as a function of disc radius and time. Internal photoevaporation is active in this simulation. Note that the relative water content is calculated as a ratio of gas surface densities, and is therefore shown as a mass fraction here. Colour coding, plotting and simulation parameters as in figure \ref{fig:gas_surface_density_and_C_to_O_no_ph}.}
    \label{fig:rel_water_with_ph}
\end{figure}

\begin{figure*}
    \centering
    \begin{minipage}{\textwidth}
        \subfigure{\includegraphics[width=0.33\textwidth]{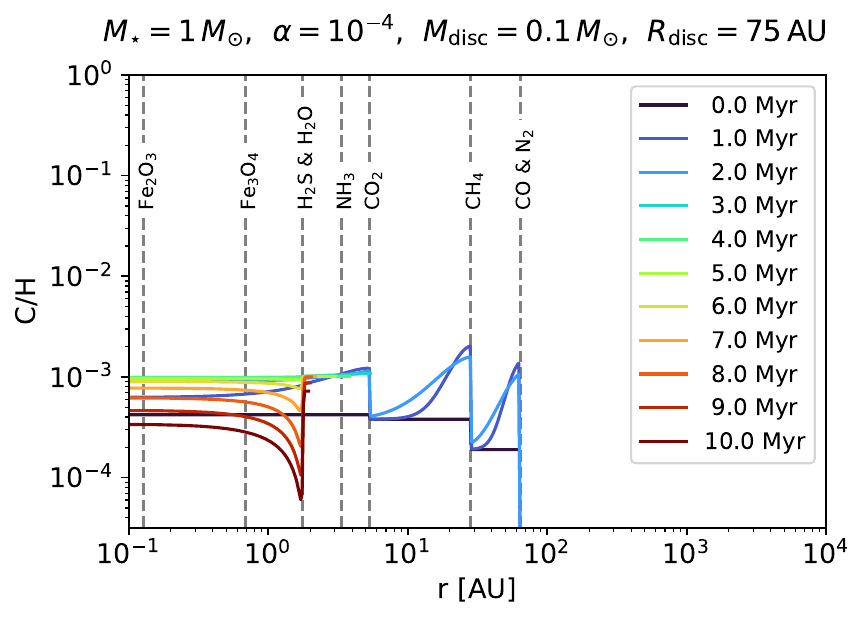}}
        \subfigure{\includegraphics[width=0.33\textwidth]{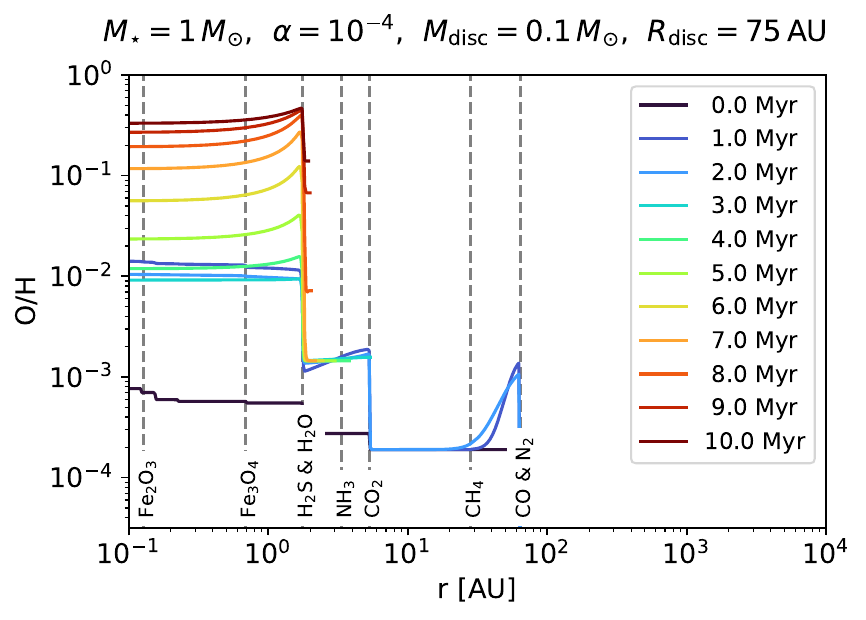}}
        \subfigure{\includegraphics[width=0.33\textwidth]{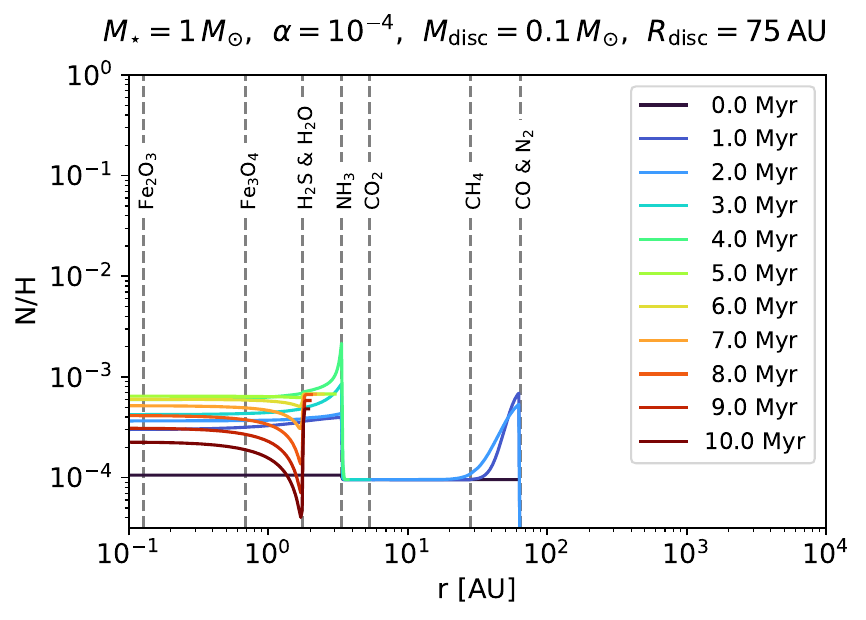}}
    \end{minipage}
    \caption{Different element ratios in the gas phase as a function of disc radius and time for a viscous disc with internal photoevaporation due to X-rays. \textbf{Left:} Carbon over hydrogen, \textbf{middle:} oxygen over hydrogen, \textbf{right:} nitrogen over hydrogen. Colour coding, plotting and simulation parameters as in figure \ref{fig:gas_surface_density_and_C_to_O_no_ph}.}
    \label{fig:CH_OH_NH_with_ph}
\end{figure*}

\begin{figure*}
    \centering
    \includegraphics[width=\textwidth]{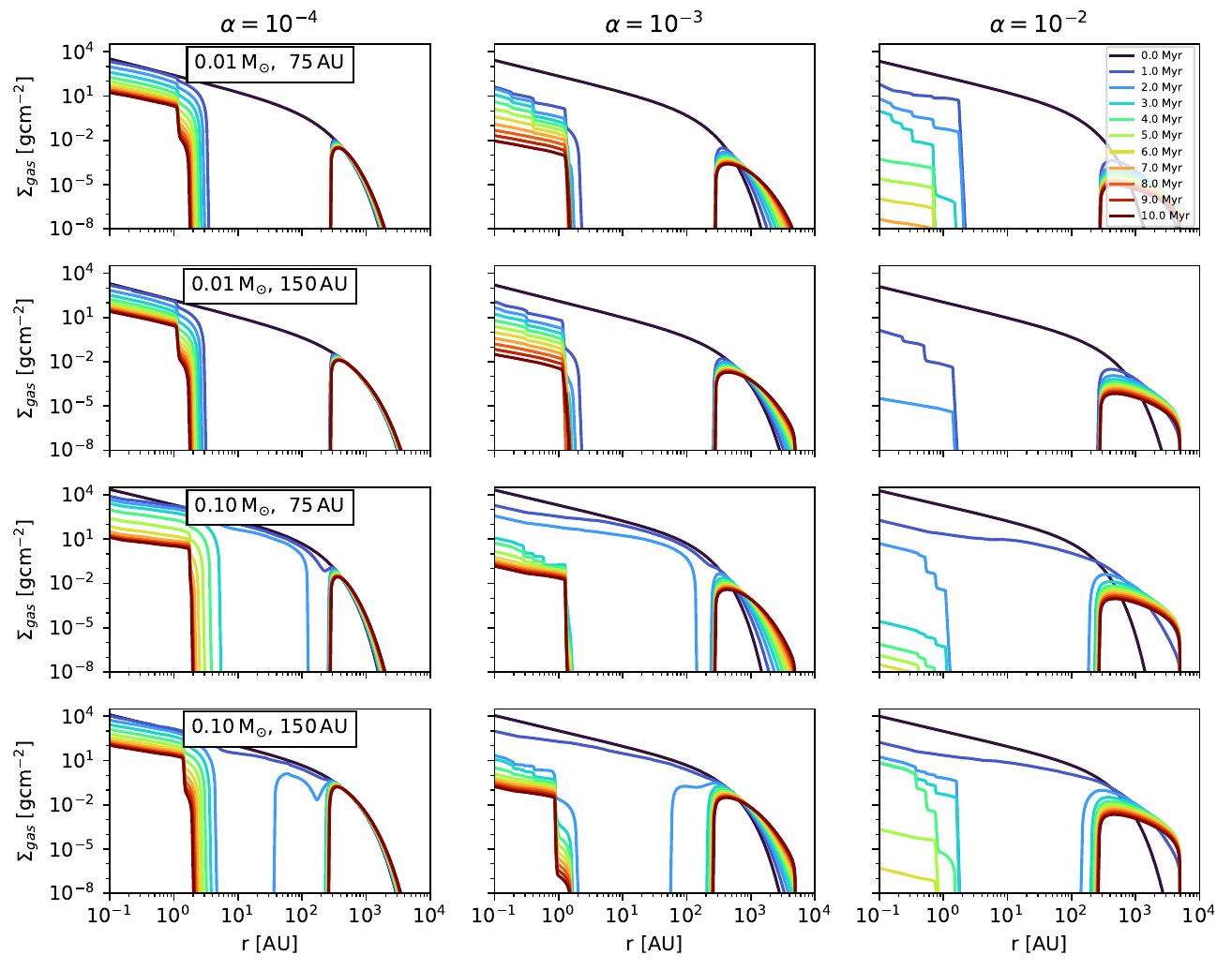}
    \caption{Gas surface density for a viscous disc with internal photoevaporation due to X-rays. The gas surface density is shown as a function of disc radius and time. A variety of parameter configurations is used for this study: Along the x-axis of this figure, $\alpha$ is varied from $\alpha = 10^{-4}$ to $\alpha = 10^{-2}$, whereas on the y-axis, different sets of disc masses and radii are used, varying in the range of $M_{\text{disc}} = 0.01 - 0.1 \, M_{\odot}$ and $R_{\text{disc}} = 75 - 150 \, \text{AU}$. Colour coding, plotting and stellar mass as in figure \ref{fig:gas_surface_density_and_C_to_O_no_ph}.}
    \label{fig:gas_surface_density_with_ph_all_par}
\end{figure*}

\begin{figure*}
    \centering
    \includegraphics[width=\textwidth]{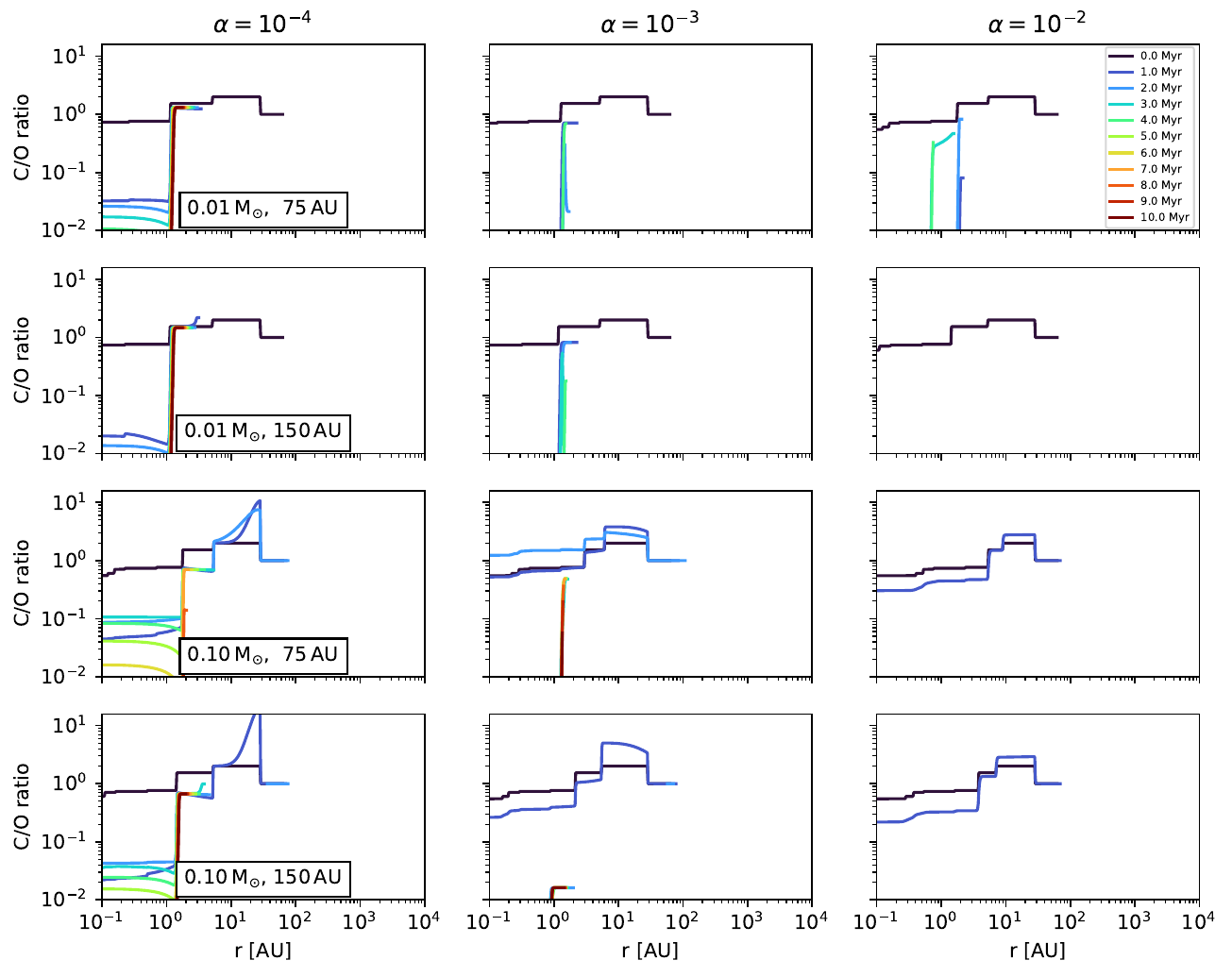}
    \caption{Gaseous C/O ratio for a viscous disc with internal photoevaporation due to X-rays. The C/O ratio is shown as a function of disc radius and time. We vary the disc parameters as in figure \ref{fig:gas_surface_density_with_ph_all_par}. Colour coding, plotting and stellar mass as in figure \ref{fig:gas_surface_density_and_C_to_O_no_ph}.}
    \label{fig:C_to_O_with_ph_all_par}
\end{figure*}

\begin{figure*}
    \centering
    \begin{minipage}{\textwidth}
        \subfigure{\includegraphics[width=0.5\textwidth]{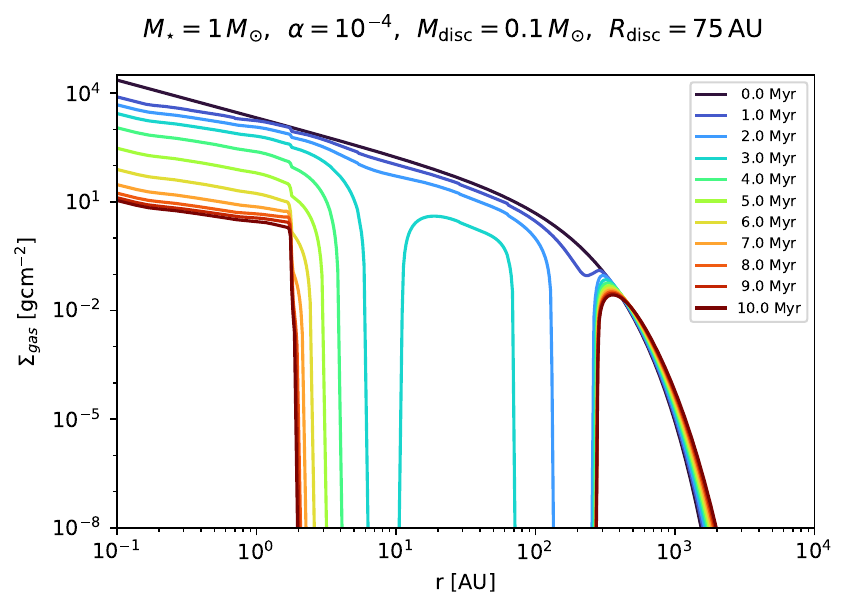}}
        \subfigure{\includegraphics[width=0.5\textwidth]{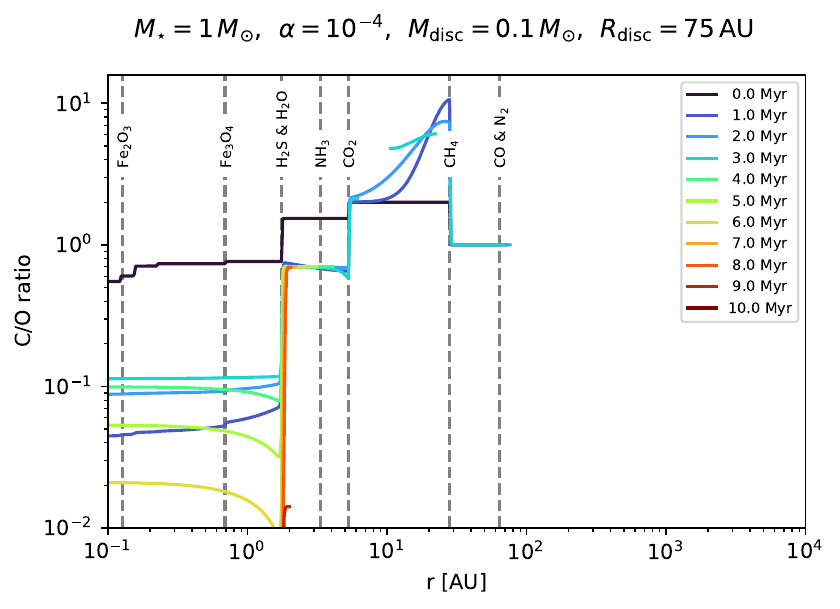}}
    \end{minipage}
    \caption{Disc evolution for a viscous disc with internal photoevaporation due to X-rays but with a smaller mass loss rate of $3.6 \cdot 10^{-8} \, M_{\odot}/\text{yr}$ instead of $3.864 \cdot 10^{-8} \, M_{\odot}/\text{yr}$. \textbf{Left:} Gas surface density as a function of disc radius and time. \textbf{Right:} Gaseous C/O ratio as a function of disc radius and time. Colour coding, plotting and simulation parameters as in figure \ref{fig:gas_surface_density_and_C_to_O_no_ph}.}
    \label{fig:gas_surface_density_and_C_to_O_with_ph_smaller}
\end{figure*}

\begin{figure*}
    \centering
    \includegraphics[width=0.95\textwidth]{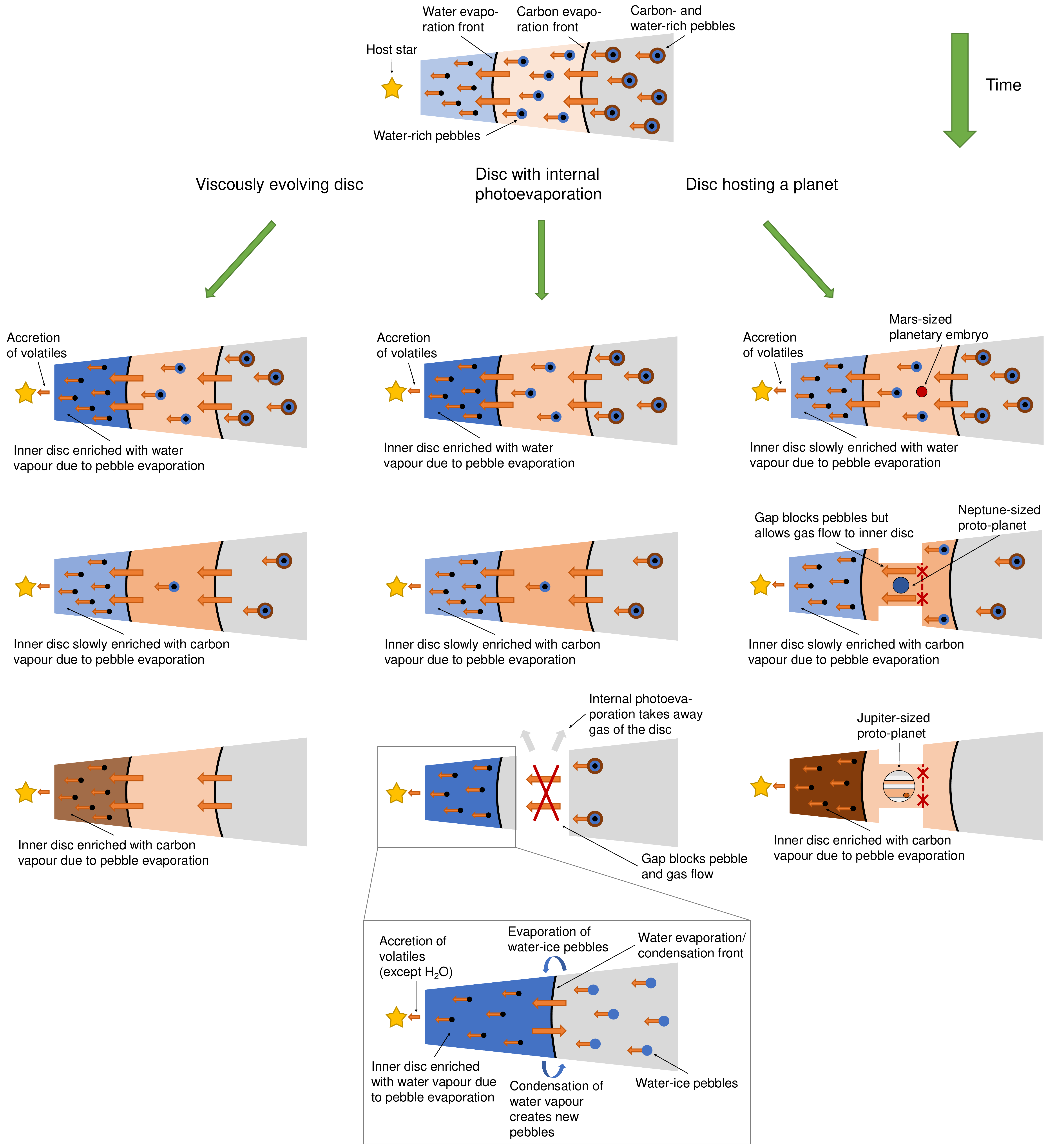}
    \caption{Phases of disc evolution, with the left column depicting a pure viscous disc, the middle column a viscous disc with internal photoevaporation, and the right column a viscous disc with a growing planet. \textbf{Top panel:} Initial state of the protoplanetary disc, with its inner part being slightly enriched with water vapour, its middle part being slightly enriched with carbon vapour and its outer regions showing no enrichment. The evaporation line of water and the combined one of all carbon-bearing molecules are marked with black, curved lines. Their curvature is the result of a temperature gradient between the mid-plane and surface layers of the disc. Pebbles are distributed evenly across the full disc, showing icy layers according to their position in it. \textbf{Left column:} During the pure viscous evolution, pebbles drifting inwards evaporate their volatile content whenever crossing an evaporation front. Due to the position of the ice lines and the fact that pebbles drift faster than gas, the inner disc is first enriched with water vapour (top panel). With time, carbon vapour arrives (middle panel) and slowly enriches the inner disc (lower panel), while the gas is accreted onto the host star. \textbf{Middle column:} With internal photoevaporation, the disc evolves similarly to the pure viscous case in the beginning (first and second panel). When photoevaporation opens a gap, gas diffusing into it is carried away by photoevaporative winds while pebbles are blocked by the gap itself (third panel). The inner disc is then enriched with water vapour again due to the water equilibrium cycle (fourth panel). All volatiles, except water, are accreted onto the host star. Water contrarily recondenses at the water-ice line and forms water pebbles that evaporate again, contributing to a full cycle of evaporation and condensation that allows the water vapour to survive for a long time. \textbf{Right column:} A planet forming between the water evaporation front and the ice lines of carbon-bearing molecules leads to the same chemical evolution of the disc as in the pure viscous case. When the planet is massive enough, it opens a gap, but contrary to the gap caused by photoevaporation, a gas flow to the inner disc is still possible. The only differences are less water enrichment in the beginning due to the planetary embryo already blocking some inflow (top panel) and a higher carbon enrichment in the late stages (lower panel).}
    \label{fig:Comic}
\end{figure*}

In this work, we discuss three different disc evolution scenarios: a purely viscously evolving disc (scenario 1), a viscous disc with additional internal photoevaporation due to X-rays (scenario 2), and a viscous disc with a giant planet forming in it (scenario 3). As this work primarily focuses on internal photoevaporation and its effects on disc evolution and composition, the main part of section \ref{sec:Results} is devoted to scenario 2. \\
In section \ref{ssec:Standard_case}, we discuss the results for our standard parameter case, where we compare a purely viscously evolving disc (scenario 1) to one with additional internal photoevaporation (scenario 2) and a viscous disc with a forming planet (scenario 3). Section \ref{ssec:Parameter_study} extends the standard parameter case to a parameter study of a disc with photoevaporation. We conclude section \ref{sec:Results} with a special case, where we compare our standard case of a disc with internal photoevaporation to a disc with a lower photoevaporative mass loss rate in section \ref{ssec:Lower_photoevaporation_rate}.

\subsection{Standard case} \label{ssec:Standard_case}
All results shown in this subsection are obtained using the standard simulation parameters as described in section \ref{ssec:Initial_conditions} and shown in table \ref{table:sim_par}.

\subsubsection{Scenario 1: Pure viscous disc} \label{sssec:Scenario_1:Pure_viscous_disc}
In the top row of figure \ref{fig:gas_surface_density_and_C_to_O_no_ph}, we show the gas surface density on the left and the gaseous C/O ratio on the right, both as a function of disc radius, for a viscous disc without internal photoevaporation. The time evolution over a span of $10 \, \text{Myr}$ is given by colour coding. The C/O ratio is calculated from number densities, normalised to solar values. \\
Our simulations confirm the expected behaviour for a viscously evolving disc with regard to the gas surface density. It decreases over time in the inner disc due to the viscous accretion of disc material onto the host star. Additionally, the outer parts of the disc are dominated by viscous spreading. Both of these processes happen relatively slowly as the alpha parameter is small, $\alpha = 10^{-4}$. \\
The initial C/O ratio shows the classical step-like behaviour \citep[e.g.][]{obergEFFECTSSNOWLINESPLANETARY2011,molliereInterpretingAtmosphericComposition2022}, where the different volatiles evaporate at different distances, resulting in a change in the C/O ratio. The positions of these evaporation fronts are determined by the disc temperature. The C/O ratio either increases at the evaporation lines of carbon-rich molecules or decreases at those of oxygen-rich molecules. \\
As the disc starts to evolve, the C/O ratio decreases rapidly in the first one million years in the inner disc, interior to the water-ice line. With time, it increases again, reaching values even higher than the initial ones at the end of the time evolution \citep[see also][]{mahCloseinIceLines2023}. This is a direct consequence of pebble drift and evaporation. Since the position of the water evaporation line is at a smaller disc radius than those of carbon-bearing molecules, pebbles from the outer regions moving inwards through the disc therefore first reach the evaporation lines of CO, CO$_2$ and CH$_4$ and evaporate their carbon content before evaporating water vapour. This means that the water vapour is formed closer to the star. In addition, pebbles move much faster than gas, water-ice pebbles thus reach the inner parts of the disc much earlier than the carbon-rich gas and evaporate their water content there. Therefore, the inner disc is enriched with water vapour first. The large amount of oxygen in water leads to a low C/O ratio. With time, more and more carbon-rich gas enters the inner disc region, elevating the C/O ratio again. At the same time, the material of the inner disc is accreted onto the host star. In the end, the C/O ratio reaches supersolar values. \\
The high C/O ratio we see in the inner disc at the end of the time evolution is a consequence of the large methane abundance in our code, see table \ref{table:molecular_species}. If methane were less abundant, the C/O ratio would not be able to increase as much as it does here and reach values larger than 1 \citep[see also][]{mahCloseinIceLines2023}. \\
Our results for a viscously evolving disc essentially reproduce studies by e.g. \cite{pisoSNOWLINELOCATIONSPROTOPLANETARY2015}, \cite{boothChemicalEnrichmentGiant2017}, \cite{bitschDryWaterWorld2021}, \cite{kalyaanLinkingOuterDisk2021}, \cite{eistrupChemicalEvolutionIces2022} or \cite{bitschEnrichingInnerDiscs2023}, who have also shown that it is important to study the chemical evolution of protoplanetary discs caused by pebble flow.

\subsubsection{Scenario 2: Viscous disc with additional internal photoevaporation} \label{sssec:Scenario_2:Viscous_disc_with_additional_internal_photoevaporation}
In a disc, where in addition to a pure viscous evolution, as in the top row of figure \ref{fig:gas_surface_density_and_C_to_O_no_ph}, internal photoevaporation is active as well, the behaviour of the gas surface density and the C/O ratio changes significantly, see middle row of figure \ref{fig:gas_surface_density_and_C_to_O_no_ph}. \\
Between $2 - 3 \, \text{Myr}$, a large gap opens in the gas surface density, spanning from $5.5 \, \text{AU}$ to $117 \, \text{AU}$. With time, its inner edge moves even further inward in the disc, which is a result of the particular shape of the gas surface density loss rate by photoevaporation (see figure \ref{fig:sig_dot}). The opened gap leads to a blocking of the pebbles from the outer disc regions. At the same time, inward-moving gas is carried away by photoevaporative winds. The gap itself is a direct result of internal photoevaporation, where gas is taken from the disc at the location where photoevaporation has the strongest impact. This is not necessarily at the position of the peak of the photoevaporative mass loss rate, as we see in our data. The left panel of the middle row of figure \ref{fig:gas_surface_density_and_C_to_O_no_ph} shows that the gap starts to open around $110 \, \text{AU}$, but the peak of the photoevaporative surface density loss rate is at $5 \, \text{AU}$, see figure \ref{fig:sig_dot}. The reason is that the disc's surface density decays exponentially in its outer regions resulting in the fact that the relative impact of photoevaporation is strongest here and not at its peak. \\
The gap has a large impact on the chemical evolution of protoplanetary discs, as can be seen from the C/O ratio (right panel of the middle row of figure \ref{fig:gas_surface_density_and_C_to_O_no_ph}). It remains subsolar in the inner disc throughout the whole evolution and never reaches supersolar values as in the case of a purely viscously evolving disc. More specifically, the C/O ratio in the case with internal photoevaporation shows similar behaviour to that of the pure viscous case in the first two million years since the gap has not opened yet. During that time, it decreases rapidly in the inner disc compared to its initial state, resulting from the evaporation of water-rich pebbles. After some time, carbon-rich gas slowly starts to enrich the inner disc as well, leading to a rise in the C/O ratio. However, before this process can lead to a supersolar C/O ratio, as in the viscous case, photoevaporation takes away too much disc material and a gap opens, as can be seen in the left panel of the middle row of figure \ref{fig:gas_surface_density_and_C_to_O_no_ph}. The pebbles from outer disc regions are blocked and gas is carried away by photoevaporative winds. Therefore, no more carbon-rich gas reaches the inner disc, the C/O ratio remains low. \\
To understand why the C/O ratio drops again after the gap-opening, we take a closer look at the water content in the inner disc. In figure \ref{fig:rel_water_with_ph}, the relative water abundance with respect to the total gas surface density is shown as a function of disc radius. We see that after an initially low abundance, it increases in the first one million years when the water vapour, evaporated from pebbles, arrives. With the arrival of carbon vapour, the water abundance decreases for the next two million years. After that, the gap is fully opened and the flux of carbon vapour is blocked, leading to an increase in the relative water abundance. The reason for this is that the gas already present in the inner disc is accreted onto the host star, with one exception. Since the water evaporation front lies within the inner disc, a water equilibrium cycle is present. Water vapour can recondense at the water-ice line and form water-ice pebbles, which in turn drift inwards and evaporate again. This enables water to stay in the inner disc for much longer than other species like CO, CO$_2$, and CH$_4$ that have their evaporation lines further out and not within the inner disc. These species are then either accreted onto the central star or removed via outward diffusion into the photoevaporation zone. The water equilibrium cycle is so effective that the relative water abundance in the inner disc increases until it makes up more than $90 \%$ of the inner disc mass at the end of its lifetime. \\
This equilibrium cycle is also the reason for the decrease in C/O in discs with internal photoevaporation after three million years, as seen in the right panel of the middle row of figure \ref{fig:gas_surface_density_and_C_to_O_no_ph}. The carbon-bearing molecules are accreted onto the host star, while water, and therefore oxygen, remains present in the inner disc for much longer due to the above-mentioned equilibrium cycle. Consequently, the C/O ratio decreases over time again. This is the exact opposite of what happens in the pure viscous case. \\
Additionally, we investigated the time evolution of the C/H, O/H and N/H abundances in the inner disc for the case with internal photoevaporation, see figure \ref{fig:CH_OH_NH_with_ph}. As expected, there is not much carbon and nitrogen but a lot of oxygen in the inner disc at the end of the disc's lifetime. This is in agreement with our study of the water content in the inner disc. For the calculation of the elemental abundance ratios, no distinction is made as to which molecule the elements belong to. Therefore, C/H and N/H decrease rapidly at the water-ice line because of the hydrogen excess due to water evaporation. Additionally, the hydrogen that is bound in water molecules remains in the inner disc due to the water equilibrium cycle, in contrast to carbon and nitrogen which are much more volatile and are being accreted onto the host star.

\subsubsection{Scenario 3: Viscous disc with a planet} \label{sssec:Scenario_3:Viscous_disc_with_a_planet}
Photoevaporation is just one possible mechanism causing gaps in protoplanetary discs, while another one is growing giant planets. The main difference is that photoevaporation takes away gas from the disc leading to a complete blocking of the pebble flow as soon as a gap is opened. Additionally, gas diffusing into the gap is carried away by photoevaporative winds. A planet on the other hand only blocks the pebbles from the outer parts, but not the gas \citep[see e.g.][]{paardekooperDustFlowGas2006,lambrechtsSeparatingGasgiantIcegiant2014,ataieeHowMuchDoes2018,bitschPebbleisolationMassScaling2018,weberCharacterizingVariableDust2018}. In this case, the gas can still move inwards through the disc, being only slightly hindered, but not blocked by the planet. \\
We study how the gas surface density and C/O ratio change for our standard parameter case when instead of internal photoevaporation, a planet is causing the gap (see bottom row of figure \ref{fig:gas_surface_density_and_C_to_O_no_ph}). A planet seed is therefore placed in the disc at $3 \, \text{AU}$, which is between the water and the CO$_2$ evaporation front in our model, at $0.05 \, \text{Myr}$. It does not migrate but grows over time due to pebble and gas accretion, where pebble accretion stops at $0.1 \, \text{Myr}$ when the planet reaches the pebble isolation mass. At this time, also the inward flux of pebbles is stopped and the inner disc cannot be reached by water-ice pebbles anymore. In the end, the planet has a mass of about $2860 \, M_{\text{Earth}}$. \\
The position of the planet is seen in the gas surface density in the left panel of the bottom row in figure \ref{fig:gas_surface_density_and_C_to_O_no_ph}. At its position, the gas surface density shows a dip, deepening with time as the planetary mass increases. \\
The C/O ratio evolves analogously to the case of a viscous disc without a planet, with the only difference that here, it is generally higher in the inner disc over the complete time evolution. The reason is that the planet reaches pebble isolation mass already after $0.1 \, \text{Myr}$, blocking all pebbles from the outer disc at this time. Since the planet is located between the water evaporation front and the ice lines of carbon-bearing molecules, water-rich pebbles have much less time than in the purely viscous disc to enrich the inner disc with water vapour before gap-opening. Therefore, the later arriving carbon-rich gas needs to balance less oxygen, leading to a higher C/O ratio. It stays continuously higher than in the pure viscous case because only water-rich pebbles are blocked. Carbon-rich pebbles have their evaporation lines further out than the planet's position. The carbon-rich gas passes the planet and enriches the inner disc. The result is a supersolar C/O ratio in the inner disc at the end of the time evolution, which is even higher than in the pure viscous case. \\
For a side-by-side comparison between the three cases discussed in this section, a viscous disc, a disc with internal photoevaporation and a disc with a forming planet, see appendix \ref{asec:Side-by-side_comparison_of_the_three_discs_presented_in_3.1}. We show there the gas surface density and the C/O ratio at $5 \, \text{Myr}$.

\subsection{Parameter study} \label{ssec:Parameter_study}
The results of our standard case in the previous section are embedded in a much larger parameter study. In this section, we investigate a range of alpha parameters, initial disc masses and radii. We use $\alpha = \{ 10^{-4}, \ 10^{-3}, \ 10^{-2} \}$, $M_{\text{disc}} = \{ 0.01 \, M_{\odot}, \ 0.1 \, M_{\odot} \}$ and $R_{\text{disc}} = \{ 75 \, \text{AU}, \ 150 \, \text{AU} \}$, the stellar mass in all simulations remains $M_{\star} = 1 \, M_{\odot}$. The results of the parameter study for the pure viscous case are shown in appendix \ref{asec:Parameter_study_of_the_viscous_case}, the ones for the viscous case with internal photoevaporation are shown here. \\
In figure \ref{fig:gas_surface_density_with_ph_all_par}, we plot the gas surface density as a function of disc radius for all the parameters mentioned above. $\alpha$ is increased along the x-axis of the grid. Each row corresponds to a specific combination of initial disc mass and radius, as depicted in the first panel of each row, respectively. Our standard case from section \ref{sssec:Scenario_2:Viscous_disc_with_additional_internal_photoevaporation} is shown in the third panel of the first column. \\
We see that the influence of $\alpha$ is strongest in the inner disc and the very outer parts. The larger $\alpha$ is, the faster the accretion of inner disc material onto the host star. For the largest value of $\alpha = 10^{-2}$, the inner disc has in all cases of initial disc mass and radius completely vanished after $10 \, \text{Myr}$. Additionally, a stronger viscous spreading in the outer disc is noticed for larger $\alpha$ values. \\
An impact of the initial disc mass on the simulations is seen in the gap-opening time. For larger initial disc masses of $M_{\text{disc}} = 0.1 \, M_{\odot}$, the gap caused by internal photoevaporation opens $1 - 2 \, \text{Myr}$ later than in the $M_{\text{disc}} = 0.01 \, M_{\odot}$ case. The reason is that for larger initial disc masses, the gas surface densities are higher and can thus withstand the photoevaporative mass loss for a longer time. \\
The effect of the initial disc radius on our simulations seems rather small because it changes the initial gas surface density and the corresponding accretion rate just by a factor of 2. However, comparing the third and the fourth panel in the first column of figure \ref{fig:gas_surface_density_with_ph_all_par}, we see that the position of gap-opening due to photoevaporation changes when using a different initial disc radius. For an initial radius of $75 \, \text{AU}$, the hole starts to open around $200 \, \text{AU}$, see the third panel of figure \ref{fig:gas_surface_density_with_ph_all_par}, whereas for an initial radius of $150 \, \text{AU}$, the position of gap-opening lies much further inward at $15 \, \text{AU}$, see the fourth panel of figure \ref{fig:gas_surface_density_with_ph_all_par}. Since the initial disc radius is defined as the location where the exponential cut-off in the initial gas surface density sets in, the two discs compared here differ in their rate of gas surface density versus photoevaporative mass loss, especially in the outer regions. This leads to photoevaporation being dominant in very different regions of those discs and thus in different gap-opening locations. \\
Figure \ref{fig:C_to_O_with_ph_all_par} shows the C/O ratio as a function of disc radius for a viscous disc with active internal photoevaporation. The parameter grid used here is the same as in figure \ref{fig:gas_surface_density_with_ph_all_par}. \\
A similar trend as in the standard case, which is depicted in the first panel of row three, is seen for a similar parameter configuration, $\alpha = 10^{-4}$, $M_{\text{disc}} = 0.1 \, M_{\odot}$ and $R_{\text{disc}} = 150 \, \text{AU}$, see first panel of row four. \\
In general, all parameter configurations show a similar behaviour, with subsolar C/O ratios in the inner disc regions, some of them being even completely depleted of carbon and oxygen, and therefore showing no C/O ratio at all. The latter is seen only in discs with larger values of $\alpha$, because in those discs, the viscous transport is so fast and dominant that the inner disc is fully accreted onto the host star, with no carbon or oxygen left to calculate a C/O ratio from. \\
There is only one configuration where we have a supersolar C/O ratio in the inner disc after $2 \, \text{Myr}$ ($\alpha = 10^{-3}$, $M_{\text{disc}} = 0.1 \, M_{\odot}$ and $R_{\text{disc}} = 75 \, \text{AU}$), which then vanishes during the rest of the disc lifetime as well. This trend is only seen in this specific parameter configuration because the $\alpha$ value is high enough to transport a lot of carbon-rich material into the inner disc before the gap opens, compared to the parameter configuration with a lower value of $\alpha$, see panel three of column one in figure \ref{fig:C_to_O_with_ph_all_par} ($\alpha = 10^{-4}$, $M_{\text{disc}} = 0.1 \, M_{\odot}$ and $R_{\text{disc}} = 75 \, \text{AU}$). This leads to an elevated amount of carbon-rich material and a therefore high C/O ratio in the inner disc during the first $2 \, \text{Myr}$. With a smaller $\alpha$ ($\alpha = 10^{-4}$), the material transport is just not fast enough to result in a supersolar C/O ratio. Furthermore, the initial disc radius plays an important role as well since it defines the location of exponential cut-off in the initial gas surface density. The disc with the supersolar C/O ratio (panel three of column two in figure \ref{fig:C_to_O_with_ph_all_par}) therefore has another rate of gas surface density versus photoevaporative mass loss than the same disc but with a larger initial disc radius ($\alpha = 10^{-3}$, $M_{\text{disc}} = 0.1 \, M_{\odot}$ and $R_{\text{disc}} = 150 \, \text{AU}$), see panel four of column two in figure \ref{fig:C_to_O_with_ph_all_par}. As stated before, these two discs then also differ in their location of gap-opening. Since the gap for the disc with $R_{\text{disc}} = 75 \, \text{AU}$ opens further out than that of the disc with $R_{\text{disc}} = 150 \, \text{AU}$, carbon-rich material can still evaporate and enrich the inner disc inside of the hole. This is not possible in the case of a larger initial disc radius because the gap starts to open in the inner disc, that is, further in than the ice lines of carbon-bearing molecules, and therefore directly hindering evaporated carbon-rich vapour from reaching the inner disc. The special behaviour of having a supersolar C/O ratio during the first $2 \, \text{Myr}$ (see panel three in column two of figure \ref{fig:C_to_O_with_ph_all_par}) of course matches the behaviour of the other panels once the gap is fully opened, reaching inner disc regions and thus hindering carbon-rich vapour from entering the inner disc. \\
We also note that in the low-viscosity case ($\alpha = 10^{-4}$), a larger initial disc mass leads to an elevated C/O ratio in the inner disc, compared to smaller initial disc masses. This is because higher initial disc masses come along with higher gas surface densities that can withstand photoevaporative mass loss for a longer time, which causes the gap to open later than in the case of a disc of lower mass. Consequently, carbon-rich material from the outer disc has more time to move towards the inner disc and enrich it.

\subsection{Lower photoevaporation rate} \label{ssec:Lower_photoevaporation_rate}
Since the X-ray luminosity for a given stellar mass is not unambiguously defined, the photoevaporative mass loss rate due to X-rays can vary. A lower limit for solar-mass stars can be read off the lower end of the error bar in figure 6 of \cite{picognaDispersalProtoplanetaryDiscs2021a}. For the study in this section, we therefore use a mass loss rate of $3.6 \cdot 10^{-8} \, M_{\odot}/\text{yr}$ instead of $3.864 \cdot 10^{-8} \, M_{\odot}/\text{yr}$ for our standard simulation to study the effects of a lower mass loss rate on our previous results. The disc parameters used here are given in table \ref{table:sim_par}. \\
In the gas surface density in the left panel of figure \ref{fig:gas_surface_density_and_C_to_O_with_ph_smaller}, we see that the gap has fully opened after $4 \, \text{Myr}$, one million years later than in the case with a slightly higher mass loss rate. For the C/O ratio in the right panel, only minor differences are noticeable. It is slightly higher than in the case of a disc with stronger photoevaporation. This is a direct consequence of the later gap-opening, ensuring that more carbon-rich gas can move into the inner disc before the gap is fully opened.

\subsection{Graphical representation of our results}
In section \ref{ssec:Standard_case}, we studied three different disc evolution scenarios. To sum up our results, we picture them in a graphical representation, see figure \ref{fig:Comic}. The left column depicts a purely viscously evolving disc (scenario 1), the middle one a viscous disc with additional internal photoevaporation due to X-rays (scenario 2), and the right one a viscous disc with a giant planet forming in it (scenario 3). \\
The protoplanetary disc is always shown as a trapezoidal shape with three areas coloured differently. This should highlight the enrichment of those disc regions with different molecules, being separated by the evaporation lines of these molecules. Those ice lines are depicted as black, curved lines, their curvature being the result of a temperature gradient between the mid-plane and surface layers of the disc \citep{dullemondPassiveIrradiatedCircumstellar2001,dullemondFlaringVsSelfshadowed2004}. As an example, we focus on the two evaporation lines that are most important for our study, namely the water-ice line and the combined evaporation front of all carbon-bearing volatiles. Pebbles, displayed as black circles, are distributed evenly across the full disc, showing icy layers according to their position. \\
All three different evolution scenarios start with the same initial condition, shown in the top panel of figure \ref{fig:Comic}. The inner disc is initially slightly enriched with water vapour, the middle part is slightly enriched with carbon vapour and the outer part is not enriched at all. Pebbles in the outer part still contain all their volatile content in ice form. \\
With time evolving continuously, pebbles and gas drift through the disc. When crossing an evaporation line, pebbles evaporate their respective volatile content, enriching the disc there. Gas reaching the very inner parts of the disc is accreted onto the host star. Depending on the scenario chosen, disc enrichment and pebble drift differ, see the lower panels of figure \ref{fig:Comic}. \\
In a purely viscously evolving disc (scenario 1), which is depicted in the left column of figure \ref{fig:Comic}, inward drifting pebbles evaporate their volatile content whenever crossing an evaporation front. Due to the position of the ice lines and the fact that pebbles drift faster than gas, the inner disc is first enriched with water vapour (top panel). With time, carbon vapour arrives (middle panel) and slowly enriches the inner disc (lower panel), while the gas is simultaneously accreted onto the host star. This can be seen in our results in section \ref{sssec:Scenario_1:Pure_viscous_disc} as well. \\
In the case of a viscous disc with additional internal photoevaporation (scenario 2), see middle column of figure \ref{fig:Comic}, the disc evolves similarly to the pure viscous case in the beginning (first and second panel). When photoevaporation opens a gap, gas diffusing into it is carried away by photoevaporative winds while pebbles are blocked by the gap itself (third panel). The inner disc is then enriched with water vapour again due to the water equilibrium cycle (fourth panel). All volatiles, except water, are accreted onto the host star. Water contrarily recondenses at the water-ice line and forms water-ice pebbles that evaporate again, contributing to a full cycle of evaporation and condensation that allows the water vapour to survive for a long time, see also our results in section \ref{sssec:Scenario_2:Viscous_disc_with_additional_internal_photoevaporation}. \\
A planet forming between the water evaporation front and the ice lines of carbon-bearing molecules (scenario 3), as shown in the right column of figure \ref{fig:Comic}, leads to the same chemical evolution of the disc as in the pure viscous case. When the planet is massive enough, it opens a gap but contrary to the gap caused by photoevaporation, a gas flow to the inner disc is still possible (middle panel). Only the pebbles are blocked once the planet reaches pebble isolation mass. The only differences to a purely viscously evolving disc (see left column of figure \ref{fig:Comic}) are less water enrichment in the beginning due to the planetary embryo already blocking some inflow (top panel) and a higher carbon enrichment in late stages (lower panel), see also our results in section \ref{sssec:Scenario_3:Viscous_disc_with_a_planet}.

\section{Discussion} \label{sec:Discussion}
\subsection{Dependence on model assumptions}
As shown in section \ref{ssec:Parameter_study}, the results of our simulations strongly depend on the chosen disc parameters. In this section, we qualitatively discuss some of the parameters that we have not investigated in detail in this work and hint at future investigations.

\subsubsection{Chemistry model} \label{sssec:Chemistry_model}
Our results clearly depend on the underlying chemistry model, as presented in table \ref{table:molecular_species}. Varying the initial composition will result in a changed significance of individual evaporation lines \citep{schneiderHowDriftingEvaporating2021,schneiderHowDriftingEvaporating2021a}. Models that include different fractions of water and carbon- and oxygen-bearing molecules will therefore show a different water content in the inner disc as well as different C/O ratios. \\
For our initial elemental abundances, we used solar values. But to be able to compare our results to observations of protoplanetary discs, it is beneficial to use observed stellar abundances since stars can differ quite a lot in their C/O ratio which is used as an initial condition (see e.g. \cite{buderGALAHSurveySecond2018} for elemental abundances and \cite{bitschInfluenceSubSupersolar2020} or \cite{cabralHowOriginStars2023} for molecular abundances). Additionally, the stellar abundance is not determined by the stellar mass but rather by the formation region of the system. This implies that to determine correctly the initial C/O ratio for individual systems, one should keep in mind that the stellar compositions are not necessarily solar, even for solar-mass stars. \\
Additionally, a different disc composition leads to a change in the mid-plane temperature. However, changing only the composition does not affect the latter very much. Stronger effects are only seen when choosing a larger disc metallicity \citep{bitschInfluenceGrainSize2021}. Since the mid-plane temperature influences the position of the evaporation lines, a change will have a significant impact on the chemical composition of the inner disc, but only if the water-ice line is shifted so far out that it does not lie in the inner disc anymore. This implies that the temperature evolution of the disc is important for the continuous water evaporation/recondensation cycle in the inner disc and would depend on the specific system.

\subsubsection{Fragmentation velocity}
How fast a pebble can become before a collision leads to fragmentation instead of coagulation is determined by the fragmentation velocity. Since collision properties of dust aggregates are not easy to quantify theoretically, fragmentation velocities are measured in laboratory experiments \citep[e.g.][]{poppeAnalogousExperimentsStickiness2000,blumGrowthMechanismsMacroscopic2008,gundlachSTICKINESSMICROMETERSIZEDWATERICE2014,musiolikContactsWaterIce2019}. Typical results range from $1 \, \text{m/s}$ to $10 \, \text{m/s}$. \\
High fragmentation velocities result in larger grains (higher Stokes numbers) and high pebble velocities. Consequently, lower fragmentation velocities lead to a decrease in the grain size and reduced pebble velocities \citep{brauerCoagulationFragmentationRadial2008,birnstielSimpleModelEvolution2012}. \\
In our simulations, we use a variable fragmentation velocity that changes its value from $1 \, \text{m/s}$ to $10 \, \text{m/s}$ at the water-ice line in the radial direction. This is motivated by the idea that water-ice grains can reach larger fragmentation velocities before they fragment compared to silicate grains \citep{gundlachSTICKINESSMICROMETERSIZEDWATERICE2014}. \\
Using instead a constant fragmentation velocity of $1 \, \text{m/s}$ throughout the whole disc would slow down all initial processes where material is transported radially inwards through the disc. A slower drift of the pebbles leads to less carbon-rich gas being evaporated and thus even less carbon vapour in the inner disc before the gap opens, compared to the case of a variable fragmentation velocity. This results in a smaller C/O ratio. As the same holds for the transport of water, its equilibrium cycle will most likely just start slower. The results for the relative water abundance and the C/O ratio in the inner disc will therefore not change considerably. \\
Increasing the overall fragmentation velocity to $10 \, \text{m/s}$ will not influence the transport from the outer to the inner disc, because the change of the fragmentation velocity just happens at the water-ice line. Therefore, the outer disc is unchanged compared to our initial assumption, and the inner disc does not change, because we only look at the gas phase \citep{bitschEnrichingInnerDiscs2023}.

\subsubsection{Stellar mass}
In this paper, all simulations have been carried out with a stellar mass of $1 \, M_{\odot}$. Changing the star's mass will also change the structure of its surrounding disc, resulting in a different initial disc mass and radius. Additionally, the disc's temperature profile changes, leading to a shift of the evaporation lines, as well as the photoevaporative mass loss rate due to X-ray radiation from the star. Since these factors cannot be considered separately due to their complex interplay, a detailed prediction of the water content and the C/O ratio in the inner disc for a non-solar mass star with active internal photoevaporation is nearly impossible. We will therefore conduct a similar, extensive study for M dwarfs in a follow-up paper. For studies of intermediate-mass stars ($1 - 3 \, M_{\odot}$) with active internal photoevaporation, see \cite{roncoPlanetFormationIntermediatemass2023}, where they investigate the effect of the latter on the disc structure without focusing on the chemical evolution.

\subsubsection{Radial position of the gap}
For the comparison between a gap caused by internal photoevaporation and one that is caused by a planet, we place a planetary seed at $3 \, \text{AU}$. The planet then forms between the water and the CO$_2$ ice line. The resulting C/O ratio in the inner disc in this case is supersolar, see section \ref{sssec:Scenario_3:Viscous_disc_with_a_planet} for more details. \\
To study the effect of the radial position of such a planetary gap, we simulate two additional scenarios. In the first one, the planetary seed is placed at $10 \, \text{AU}$, between the CO$_2$ and the CH$_4$ ice line. This results in a visible gap in the protoplanetary disc at the position of the planet. The C/O ratio in the inner disc is supersolar, as in our initial case. In the second scenario, the planetary seed is placed at an even larger disc radius of $50 \, \text{AU}$, which is between the CH$_4$ and the CO ice line. In this case, no visible gap opens up in the disc since the planet never reaches the pebble isolation mass. The resulting gas surface density and the C/O ratio show no significant deviation from the gas surface density and the C/O ratio of a purely viscously behaving disc without a planet, where the latter is shown in the top row of figure \ref{fig:gas_surface_density_and_C_to_O_no_ph}. In the end, the C/O ratio of the second scenario reaches supersolar values as well. \\
This simple comparison shows that the C/O ratio in the inner disc is independent of the planetary location. It is supersolar for all cases studied here. The results of this comparison of course strongly depend on the underlying chemistry model. For a further discussion of the latter, see section \ref{sssec:Chemistry_model}.

\subsection{Implications of our results}
\subsubsection{Disc observations}
Our results clearly indicate that gaps caused by planets and gaps caused by internal photoevaporation differ strongly in how they affect the protoplanetary disc structure and composition. Gaps caused by planets only block the pebble flow from the outer parts once the planet reaches pebble isolation mass but not the gas. Photoevaporative gaps on the other hand block not only the pebble flow but also lead to gas diffusing into them being carried away by photoevaporative winds. These mechanisms leave no room for further enrichment of the inner disc once a gap is fully opened. This results in very different chemical compositions of the inner disc regions, where discs hosting planets can have supersolar C/O ratios, whereas discs with active internal photoevaporation have difficulties reaching supersolar C/O ratios, see sections \ref{sssec:Scenario_2:Viscous_disc_with_additional_internal_photoevaporation} and \ref{sssec:Scenario_3:Viscous_disc_with_a_planet}. As a consequence, we should be able to distinguish observationally between these two scenarios when measuring the C/O ratio. This implies that we can infer the root cause of a gap structure when observing protoplanetary discs. Since these gap structures are equally seen in the gas and dust phases in our simulations, observations are not limited to one or the other. \\
\cite{banzattiJWSTRevealsExcess2023} find in their study of the spectra of four discs observed with the \ac{JWST} \ac{MIRI} that the water abundance is mainly dependent on the outer disc radius. They show that the analysed spectra of the two compact discs ($10 - 20 \, \text{AU}$) they studied hint towards a water excess compared to the two large discs with multiple gaps ($100 - 150 \, \text{AU}$). Another \ac{JWST} observation shows a big water reservoir in a large disc with substructures \citep{gasmanMINDSAbundantWater2023}. This may hint towards the fact that gap structures play a more important role in creating water reservoirs than the disc radius. In agreement with the observation by \cite{gasmanMINDSAbundantWater2023}, we find that the disc radius does not play a major role in the time evolution of the gas surface density and the C/O ratio and therefore the water content of the inner disc, see section \ref{ssec:Parameter_study}. Far more important than the disc radius is the presence of gaps in protoplanetary discs and whether they are caused by planets or photoevaporation, see sections \ref{sssec:Scenario_2:Viscous_disc_with_additional_internal_photoevaporation} and \ref{sssec:Scenario_3:Viscous_disc_with_a_planet}. Our results point to a planetary origin of the gaps in the two extended discs studied by \cite{banzattiJWSTRevealsExcess2023}. This stems from the fact that we see a water excess in our results only in discs where the substructures are caused by internal photoevaporation. However, the two extended discs studied by \cite{banzattiJWSTRevealsExcess2023}  show no water excess. \\
\cite{grantMINDSDetection132023} have shown that the disc around the T Tauri star GW Lup is very abundant in CO$_2$, using observations from the \ac{JWST}-\ac{MIRI} programme: \ac{MINDS}. As a possible explanation, they proposed among other scenarios an inner cavity between the H$_2$O and CO$_2$ ice lines. With our model, we are now able to identify the cause of this possible gap. Using our arguments above, a high abundance in CO$_2$ points strongly towards a planet in this inner cavity, rather than the gap being formed by internal photoevaporation. Although GW Lup has a stellar mass of $M_{\star} = 0.46 \, M_{\odot}$ \citep{alcalaXshooterSpectroscopyYoung2017,andrewsDiskSubstructuresHigh2018}, the main trend of our study for solar-mass stars, namely a clear compositional difference of the inner disc depending on which scenario is the main cause for the gap-opening, will still be valid in this mass regime. The reason is that here the exact positions of the ice lines do not matter, but instead their relative locations. \\
Measured CO$_2$ abundances of \cite{grantMINDSDetection132023} are higher than what we find is possible to achieve in discs with active internal photoevaporation. However, GW Lup has a lower stellar mass than the stars in our simulations. Studies of \cite{mahCloseinIceLines2023} for discs without active internal photoevaporation have shown that less massive stars achieve supersolar C/O ratios in the inner disc faster since all of the ice lines are closer to the host star. Furthermore, the work of \cite{picognaDispersalProtoplanetaryDiscs2021a} for discs with active internal photoevaporation points out that inner disc lifetimes increase for less massive stars, indicating that more carbon-rich gas can move to the inner disc. The combination of those two factors might then allow for higher C/O ratios in photoevaporative discs around low-mass stars compared to those around solar-like stars. \\
The lifetimes of the discs studied by \cite{banzattiJWSTRevealsExcess2023} and \cite{grantMINDSDetection132023} are shorter than the timescale in our model that is necessary to achieve a supersolar C/O ratio in the inner disc. However, our standard disc, with which we compare here, is modelled using a small value for the turbulence parameter $\alpha$ and has therefore a low viscosity. Changing the disc's viscosity directly impacts the timescale for the evaporation of molecules and the subsequent enrichment of the gas with vapour. Larger viscosities result in smaller formation timescales of supersolar C/O ratios \citep[see figure \ref{fig:C_to_O_no_ph_all_par} and also][]{mahCloseinIceLines2023}, making them again comparable to the disc lifetimes in \cite{banzattiJWSTRevealsExcess2023} and \cite{grantMINDSDetection132023}.

\subsubsection{Planetary compositions}
From the work done in this paper, we can infer implications for the atmospheric composition of planets forming in discs with active internal photoevaporation. \\
If a planet forms in the inner disc, it will accrete its atmosphere from the volatile content present there. One would naively assume that such a planet in our model will have a low atmospheric C/O ratio and a high water abundance. However, we have to distinguish between different formation timescales. \\
On the one hand, when a planet forms early on, it might stay in the inner disc for a longer time. Due to the existence of the water equilibrium cycle and the water thus being present there for a long time as well, the planet could accrete a high amount of water, possibly forming a very water-rich planet. This is supported by observational data from PDS 70 showing that there are indeed systems with water being abundant in the inner disc \citep{perottiWaterTerrestrialPlanetforming2023}. However, the presence of a planet in the inner disc will most likely perturb the water equilibrium cycle. With the planet accreting part of the water vapour, less water remains to go through the equilibrium cycle, thus reducing it. In this case, the planet will most likely end up with a less water-abundant atmosphere than in the previously discussed case. On the other hand, a late-forming planet might therefore have better chances to accrete a water-rich atmosphere. The reason is that the water equilibrium cycle has not been perturbed yet, giving it more time to keep the water in the inner disc. To conclude, the composition of planetary atmospheres is strongly dependent on the point in time when the planet starts to accrete gas efficiently. \\
In addition to the formation timescale, the formation position and migration of a planet are additional factors that play a major role in the planet's ability to accrete certain molecular species. This makes it overall very difficult to precisely predict the planet's final atmospheric composition in a disc with active internal photoevaporation.

\subsubsection{Debris discs}
Our simulations show that a disc with active internal photoevaporation still has material left in the region outside of the gap carved by photoevaporative winds at the end of its lifetime. However, the total remaining dust mass of $1.3 \, M_{\text{Earth}}$ in the outer disc, when converted to planetesimals, is not enough to account for debris discs, for which several tens of Earth masses are needed.

\section{Summary and conclusions} \label{sec:Summary_and_conclusions}
We performed 1D semi-analytical simulations of protoplanetary discs with active internal photoevaporation due to X-rays for solar-like stars. The model we used includes pebble drift and evaporation. Combining a pebble drift and evaporation model with internal photoevaporation opens up new perspectives for understanding the composition of inner discs. We compare our results to discs that evolved purely viscously without photoevaporation. \\
Our results clearly indicate that internal photoevaporation has significant effects on the evolution of protoplanetary discs and their chemical composition. The consequential opening of gaps strongly influences the distribution of gas and dust in such discs. Due to the diffusion of gas into these gaps ---which is then carried away by photoevaporative winds--- and the consecutive blocking of pebbles by the gaps themselves, it is impossible to have solar or even supersolar C/O ratios in the inner disc. Additionally, the present equilibrium water cycle allows the formation of very water-rich inner disc regions. Potential planets forming in these areas could therefore easily have large water abundances in their atmospheres. \\
The above-mentioned gaps due to internal photoevaporation are very different from those caused by planets forming in protoplanetary discs. The latter block only the pebble flow and not the gas, leading to a disc evolution similar to that of a pure viscous disc and resulting in a supersolar C/O ratio for the inner disc region. Future observations could therefore easily infer the root cause of a gap structure when observing the C/O ratio in the inner regions of protoplanetary discs with gaps. As these structures are equally seen in the gas and dust phases in our simulations, observations are not limited to one or the other. \\
We conclude that future simulations designed to study the evolution of protoplanetary discs should never neglect internal photoevaporation, as it appears to have a significant impact on the evolution and chemical composition of the discs, and therefore on the composition of the planets forming in them.

\section*{Acknowledgements}
We thank Giovanni Picogna, Remo Burn, Manuel Güdel, James Owen and Sho Shibata for helpful discussions. \\
J.L.L. and B.B. acknowledge the support of the European Research Council (ERC Starting Grant 757448-PAMDORA). Th.H. acknowledges the support of the ERC Origins grant number 832428.

\bibliographystyle{aa}
\bibliography{Paper_1}

\begin{appendix}
\section{Side-by-side comparison of the three discs presented in \ref{ssec:Standard_case}} \label{asec:Side-by-side_comparison_of_the_three_discs_presented_in_3.1}
\begin{figure}[h]
    \centering
    \begin{minipage}{\textwidth}
        \subfigure{\includegraphics[width=0.5\textwidth]{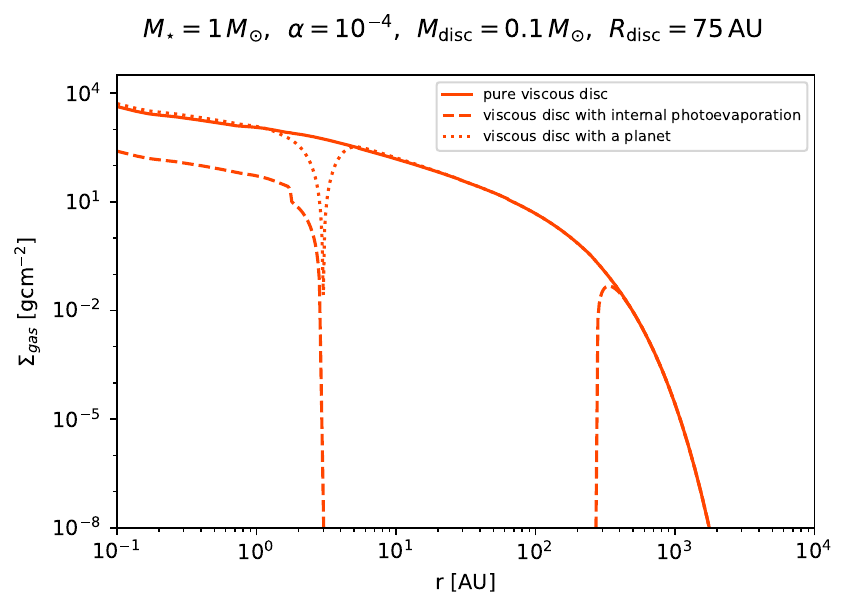}}
    \end{minipage}
    \begin{minipage}{\textwidth}
        \subfigure{\includegraphics[width=0.5\textwidth]{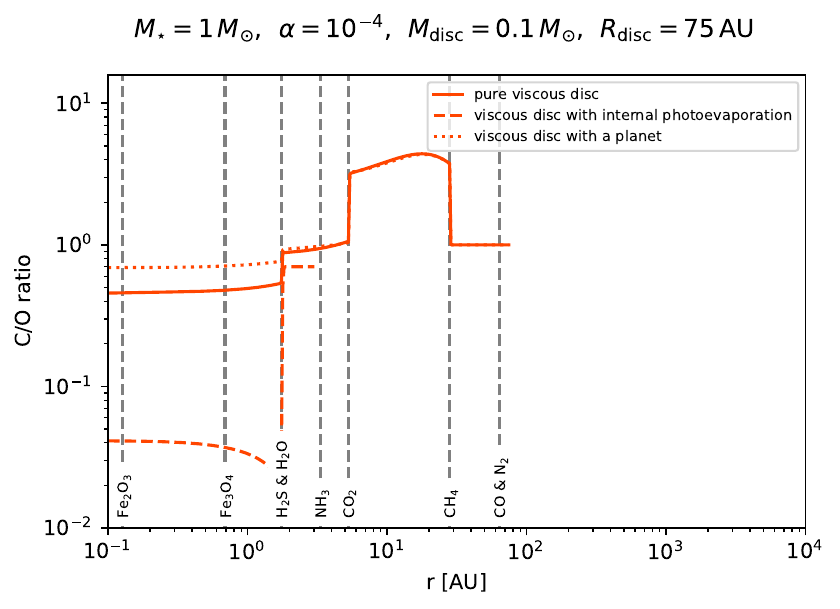}}
    \end{minipage}
    \caption{State of the disc at $5 \, \text{Myr}$ for the three different scenarios studied in this paper: a viscous disc (solid line), a viscous disc with internal photoevaporation due to X-rays (dashed line), and a viscous disc with a planet sitting at $3 \, \text{AU}$ (dotted line). \textbf{Top:} Gas surface density as a function of disc radius. \textbf{Bottom:} Gaseous C/O ratio as a function of disc radius. Plotting and simulation parameters as in figure \ref{fig:gas_surface_density_and_C_to_O_no_ph}.}
    \label{fig:gas_surface_density_and_C_to_O_comparison}
\end{figure}

Figure \ref{fig:gas_surface_density_and_C_to_O_comparison} shows a side-by-side comparison of the three disc versions that are discussed in detail in section \ref{ssec:Standard_case}. The solid line represents the purely viscous disc, the dashed line the viscous disc with internal photoevaporation via X-rays and the dotted line the viscous disc with a forming planet. In the left and right panels, the gas surface density and the C/O ratio are plotted for the time of $5 \, \text{Myr}$, respectively.
\FloatBarrier

\FloatBarrier
\section{Parameter study of the viscous case} \label{asec:Parameter_study_of_the_viscous_case}
\begin{figure*}
    \centering
    \includegraphics[width=\textwidth]{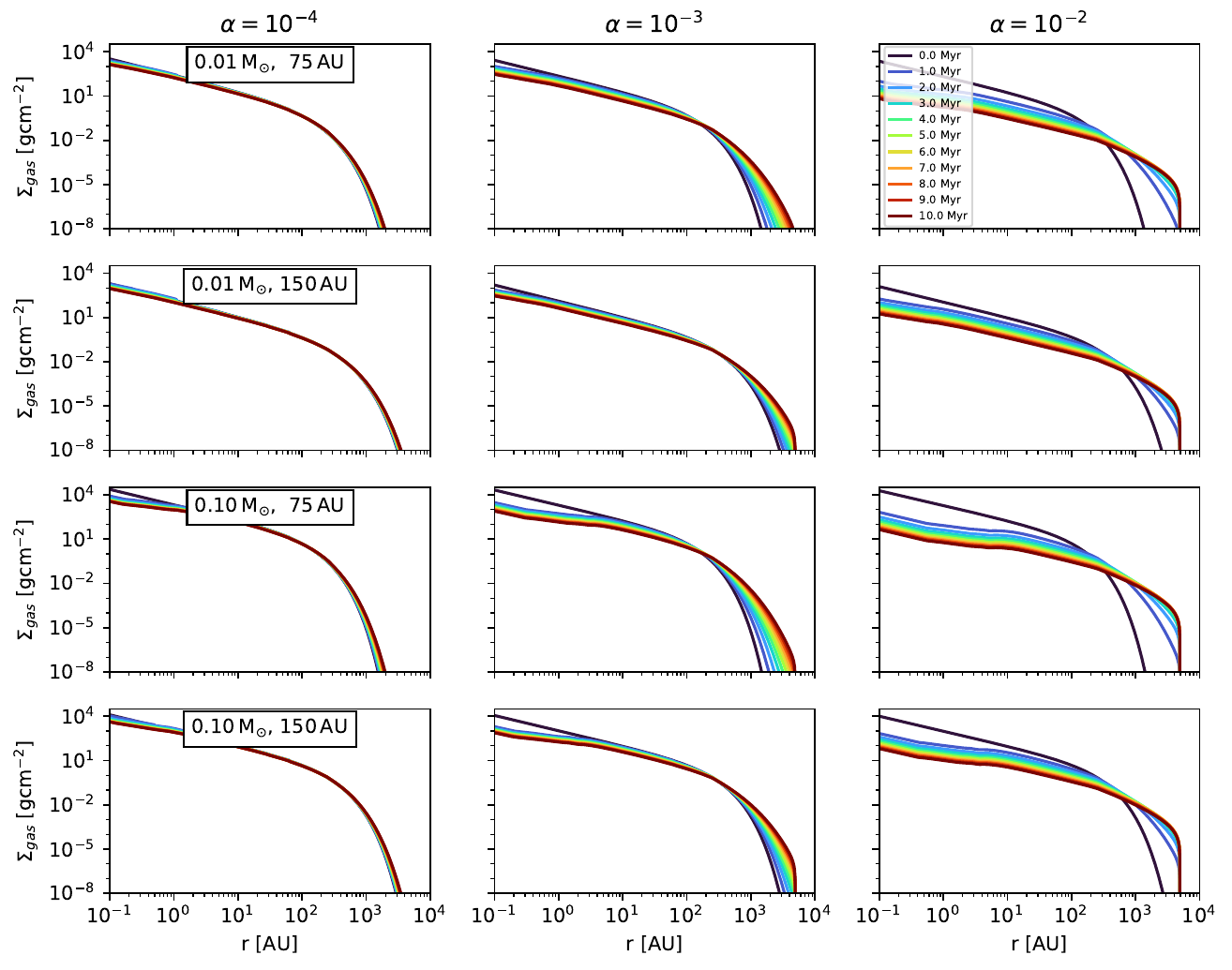}
    \caption{Gas surface density for a viscous disc without internal photoevaporation. The gas surface density is shown as a function of disc radius and time. We vary the disc parameters as in figure \ref{fig:gas_surface_density_with_ph_all_par}. Colour coding, plotting and stellar mass as in figure \ref{fig:gas_surface_density_and_C_to_O_no_ph}.}
    \label{fig:gas_surface_density_no_ph_all_par}
\end{figure*}

\begin{figure*}
    \centering
    \includegraphics[width=\textwidth]{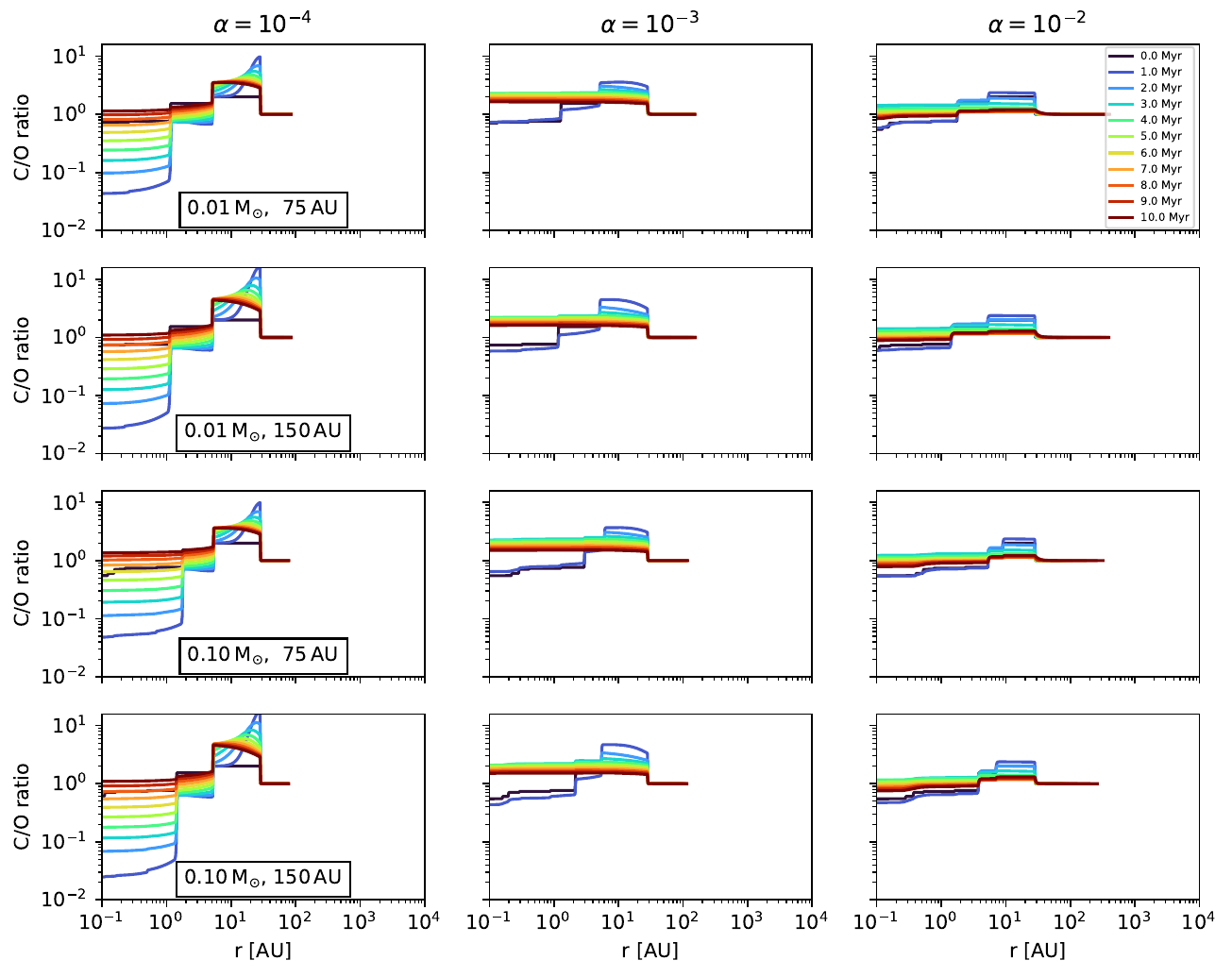}
    \caption{Gaseous C/O ratio for a viscous disc without internal photoevaporation. The C/O ratio is shown as a function of disc radius and time. We vary the disc parameters as in figure \ref{fig:gas_surface_density_with_ph_all_par}. Colour coding, plotting and stellar mass as in figure \ref{fig:gas_surface_density_and_C_to_O_no_ph}.}
    \label{fig:C_to_O_no_ph_all_par}
\end{figure*}

We present here the results of a parameter study of a purely viscously evolving disc. An analogous analysis for a viscous disc with additional internal photoevaporation is done in section \ref{ssec:Parameter_study}. Investigated here is a range of alpha parameters, initial disc masses and radii. We use $\alpha = \{ 10^{-4}, \ 10^{-3}, \ 10^{-2} \}$, $M_{\text{disc}} = \{ 0.01 \, M_{\odot}, \ 0.1 \, M_{\odot} \}$ and $R_{\text{disc}} = \{ 75 \, \text{AU}, \ 150 \, \text{AU} \}$, the stellar mass in all simulations remains $M_{\star} = 1 \, M_{\odot}$. \\
In figure \ref{fig:gas_surface_density_no_ph_all_par}, we plot the gas surface density as a function of disc radius for all the parameters mentioned above. $\alpha$ is increased along the x-axis of the grid. Each row corresponds to a specific combination of initial disc mass and radius, as depicted in the first panel of each row, respectively. Our standard parameter case, as presented in section \ref{sssec:Scenario_1:Pure_viscous_disc}, is shown in the third panel of the first column. \\
We note that the largest visible difference between the plots is seen with varying $\alpha$. Its influence is strongest in the inner disc and the very outer parts. The larger $\alpha$ is, the faster the accretion of inner disc material onto the host star. Additionally, a stronger viscous spreading in the outer disc is seen for bigger values of $\alpha$. \\
An impact of the initial disc mass on the simulations is seen in the accretion of inner disc material onto the host star. For larger initial disc masses of $M_{\text{disc}} = 0.1 \, M_{\odot}$, disc material seems to be accreted slightly faster than in the $M_{\text{disc}} = 0.01 \, M_{\odot}$ case. The reason is that the accretion rate correlates with the initial disc mass. \\
The effect of the initial disc radius on our simulations seems rather small because it changes the initial gas surface density and the corresponding accretion rate just by a factor of 2. However, comparing the first with the second and the third with the fourth row of figure \ref{fig:gas_surface_density_no_ph_all_par}, we see a slight difference in the viscous spreading in the outer disc with varying initial disc radius. For larger initial radii, the viscous spreading is less dominant. This is because the initial disc radius is defined as the location where the exponential cut-off in the initial gas surface density sets in. Discs with larger initial radii therefore have their cut-off further out in the disc, and can therefore better balance the effect of viscous spreading in the outer parts due to more gas being present, leading to less viscous spreading. \\
Figure \ref{fig:C_to_O_no_ph_all_par} shows the C/O ratio as a function of disc radius for a disc that evolves purely viscously. The parameter grid used here is the same as in figure \ref{fig:gas_surface_density_no_ph_all_par}. \\
We see again that varying the $\alpha$ parameter has the largest visible effect on our results. For $\alpha = 10^{-4}$, the C/O ratio behaves similarly to our standard case, which is depicted in the third panel of the first column (see also section \ref{sssec:Scenario_1:Pure_viscous_disc}). However, for larger values of $\alpha$, the initial drop of the C/O ratio is not as high as in the low-viscosity case. Additionally, it increases very rapidly to supersolar values from one time step to the next, only to decrease again, but very slowly this time, until the end of the time evolution. As in the low-viscosity case, pebbles drift faster than gas and therefore, water-rich pebbles reach the inner disc first before carbon-rich gas arrives. They evaporate their water content, enriching the inner disc with oxygen, which leads to an initial decrease in C/O. However, gas is transported faster through the disc due to the higher viscosity. Carbon-rich gas therefore arrives only shortly after, the decrease is not as big as in the low-viscosity case, meaning less water vapour needs to be balanced. As more and more carbon-rich gas arrives quickly, the C/O ratio increases. With still evaporating water-rich pebbles, it balances out over time but stays supersolar. \\
Varying the initial disc mass and radius seems to have only a minor influence on the C/O ratio in the inner disc. In general, all parameter configurations show similar behaviour, with supersolar C/O ratios in the inner disc regions at the end of the time evolution.
\end{appendix}

\end{document}